\documentclass[12pt,a4paper]{article}
\usepackage{latexsym,amssymb,epsfig}

\newcommand{\bm}[1]{\mbox{\boldmath$#1$}}
\def\0{\phantom{0}}
\newfont{\tnf}{cmbx12 at 18pt}
\begin{document}
\begin{center}


{\tnf Vapor-liquid equilibria simulation\\
and an equation of state contribution \\
for dipole-quadrupole interactions}\\

\bigskip
\bigskip

\renewcommand{\thefootnote}{\fnsymbol{footnote}}
Jadran Vrabec$^1$, Joachim Gross$^2$\footnote[1]{corresponding author, phone: +31-15-278 6658, fax: +31-15-278 2460, e-mail: j.gross@tudelft.nl}\\

\textit{$^1$ Institut f\"ur Technische Thermodynamik und Thermische Verfahrenstechnik, \\
Universit\"at Stuttgart, Pfaffenwaldring 9, 70550 Stuttgart, Germany \\}
\textit{$^2$  Engineering Thermodynamics, Delft University of Technology, Leeghwaterstraat 44, 2628 CA Delft, The Netherlands\\}

\end{center}
\clearpage

\section*{Abstract}
\baselineskip24pt
A systematic investigation on vapor-liquid equilibria (VLE) of dipolar and quadrupolar
fluids is carried out by molecular simulation to develop a new 
Helmholtz energy contribution for equations of state (EOS). Twelve two-center  
Lennard-Jones plus point dipole and point quadrupole model fluids (2CLJDQ) are studied
for different reduced dipolar 
moments $\mu^{*2}=6$, 12, reduced quadrupolar moments $Q^{*2}=2$, 4 and reduced
elongations $L^*=0$, 0.505, 1. Temperatures cover a wide range from about 55 \% to
about 95 \% of the critical temperature of each fluid. The 
$N\!pT$ + test particle method is used for the calculation of vapor pressure,
saturated densities and saturated enthalpies.  
Critical data and the acentric factor are obtained from fits to the simulation data.
On the basis of this data,  
an EOS contribution for the dipole-quadrupole cross-interactions of non-spherical
molecules is developed. The expression is based on a third-order
perturbation theory and the model constants are adjusted to the present 2CLJDQ 
simulation results.
When applied to mixtures, the model is found to be in excellent
agreement to results from simulation and experiment. The new EOS contribution is also compatible with
segment-based EOS, like the various forms of the Statistical Associating Fluid Theory (SAFT) EOS.

\newpage

\section{Introduction}
\baselineskip24pt
Knowledge of thermodynamic properties, and in particular vapor-liquid equilibria (VLE), is important for many problems in science and engineering. 
Among the state-of-the-art thermodynamic models, molecular based approaches have
the highest potential to yield significant improvements compared to existing
phenomenological approaches, especially in terms of predictive power. 
 
For the two-center Lennard-Jones plus point dipole (2CLJD) fluid and its quadrupolar pendant (2CLJQ), systematic studies of VLE were carried out in previous work \cite{stoll4549,stoll179}. 
The VLE results were correlated as functions of the model parameters, which were used to determine the model parameters for 78 real fluids \cite{stoll03,vrabec105}. 
It has been shown that these molecular models can successfully be applied for the description of VLE for binary and ternary mixtures \cite{stoll000,vrabec05,stolldiss}. 
The simulation data from the two systematic studies \cite{stoll4549,stoll179} were subsequently used to construct equation of state (EOS) contributions due to  
dipolar \cite{gross06} and quadrupolar \cite{gross05} interactions.

As numerous real fluids, e.g., carbon monoxide or refrigerants like R115 (CF$_3$-CF$_2$Cl) are both dipolar and quadrupolar, 
it is valuable to study multipolar model fluids. In case of mixtures, containing dipolar components and quadrupolar components, the polar cross-interaction 
also plays a significant role.  
However, only few studies on VLE of fluids that are both dipolar and quadrupolar are available in the literature.  
Without reference to real fluids, Dubey and O'Shea \cite{dubey9421} studied VLE of one-center Lennard-Jones plus dipole and
quadrupole model fluids.

Molecular theories for multipolar compounds were developed by Stell et al.
\cite{stell1,stell2}, based on a perturbation theory around a non-polar reference fluid. A
hard-sphere reference was considered by Rushbrooke et al. \cite{rush1} and later by
Henderson et al. \cite{hen1}. The perturbation theory for mixtures was worked out by
Gubbins and Twu \cite{gub1} considering point-electrostatic multipoles with the spherical
Lennard-Jones reference fluid. The perturbation
terms were initially parameterized to molecular dynamics results and this
parameterization was subsequently improved by Luckas et al. \cite{luc1}. The
perturbation terms were in detail reported by Moser et al. \cite{mos1} and Shukla et
al. \cite{shu1} and good results were found upon applying that theory to real
mixtures \cite{shu2,shu3}. Boublik studied the effect of the non-spherical molecular shape
on multipolar interactions, considering the pair correlation function of a
Gaussian overlap fluid \cite{bou1}.

An equation of state for the 2CLJD and the 2CLJQ fluid were developed by Saager and Fischer based on molecular
simulations. A fixed elongation ($L^*=0.505$) was thereby considered and a dipole-dipole term and a
quadrupole-quadrupole term were obtained by fitting empirical expressions to the simulation data \cite{saager9127,saager1992}.
Dipole-quadrupole cross-interactions were subsequently treated with an effective one-fluid dipole moment which contains
a contribution due to the quadrupole moments of the mixture \cite{weingerl202}. Similarily an effective one-fluid
quadrupole moment was defined where the dipole moments of components in the mixture contributed according to an empirical
relation that was parameterized to simultion data.

Due to the success of the Statistical Associating Fluid Theory (SAFT) EOS \cite{chapman88,wertheim8419,wertheim8435,wertheim86459,wertheim86477} in various variants
\cite{bla1,bla2,pam1,gil1,dav1,gross00}, the multipolar terms developed earlier for spherical fluids, were recently applied in combination
with different SAFT models \cite{gross01,kra1,jog1,jog2,tum1,li1,liu1,kar1,kar2}. Zhao and McCabe used an integral-equation
approach with the mean spherical approximation (MSA) closure in combination with
the SAFT-VR EOS \cite{zha1}. As structural information is available from the MSA term,
the orientation of dipoles can be accounted for. 

The authors of the present
study have recently developed expressions that account for the non-spherical
shape of fluids and found good results also for mixtures of 2CLJQ and 2CLJD fluids
as well as for mixtures of real substances \cite{gross06,gross05}. Furthermore, Kleiner and Gross \cite{kle1}
accounted for the polarizability of fluids and for the appropriate
induction effects of dipoles in applying a renormalization scheme proposed by
Wertheim \cite{wer1,wer2,wer3,gra1}. A combination of the perturbation theory
of Moser et al. \cite{mos1} and of Shukla et al.  \cite{shu1}, where the full
vectorial and tensorial information of multipoles is utilized, with the approach
of our earlier work \cite{gross06,gross05} is presented by Leonhard et al. \cite{leon07}.

In this work, VLE of two-center Lennard-Jones model fluids which are both
dipolar and quadrupolar (2CLJDQ) are systematically studied over a wide range
of model parameters and temperatures. In reduced units, the 2CLJDQ model class
has three molecular parameters that can be varied: elongation, dipole moment and
quadrupole moment. Twelve different 2CLJDQ model fluids are covered here.
Using these results, an EOS contribution for the dipole-quadrupole
cross-interactions of non-spherical molecules is developed based on a third-order
perturbation theory. Model constants are thereby adjusted to the simulation
results of the 2CLJDQ fluid. Finally, the EOS contribution is compared to experimental 
and simulation data for five real binary mixtures.

\section{Molecular model}
The two-center Lennard-Jones plus point dipole and point quadru\-pole fluids (2CLJDQ) is here investigated. 
The pair potential of this model class is composed of two identical Lennard-Jones sites a distance $L$ apart (2CLJ) plus a point dipole of moment $\mu$ and a point quadrupole of moment $Q$ placed in the geometric center of the molecule. 
Both polarities are aligned along the molecular axis. The full potential writes as 
\begin{eqnarray}
u_{\rm 2CLJDQ}(\bm{r}_{ij},\bm{\omega}_i,\bm{\omega}_j,L,\mu,Q,\sigma,\epsilon) &=& u_{\rm2CLJ}(\bm{r}_{ij},\bm{\omega}_i,\bm{\omega}_j,L,\sigma,\epsilon)+u_{\rm DD}(\bm{r}_{ij},\bm{\omega}_i,\bm{\omega}_j,\mu)\nonumber\\
                                                                &+& u_{\rm QQ}(\bm{r}_{ij},\bm{\omega}_i,\bm{\omega}_j,Q)+u_{\rm DQ}(\bm{r}_{ij},\bm{\omega}_i,\bm{\omega}_j,\mu,Q), 
\label{fullpotential}
\end{eqnarray}
where
\begin{eqnarray}
u_{\rm 2CLJ}(\bm{r}_{ij},\bm{\omega}_i,\bm{\omega}_j,L,\sigma,\epsilon) &=& \sum_{a=1}^{2} \sum_{b=1}^{2} 4\epsilon
\left[\left( \frac{\sigma}{r_{ab}} \right)^{12} - \left( \frac{\sigma}{r_{ab}} \right)^6 \right], \0\0\0\0\0 
\end{eqnarray}
is the Lennard-Jones part. Herein, ${\bm r}_{ij}$ is the center-center distance vector of two molecules
$i$ and $j$, $r_{ab}$ is one of the four Lennard-Jones site-site distances, $a$ counts the two sites of molecule $i$, $b$ counts those of molecule $j$. The vectors ${\bm
\omega}_i$ and ${\bm \omega}_j$ represent the orientations of the two molecules. There are three polar contributions, cf.~Allen and Tildesley \cite{allen87}: Firstly, the
interaction between dipoles is present 
\begin{eqnarray}
u_{\rm DD}(\bm{r}_{ij},\bm{\omega}_i,\bm{\omega}_j,\mu) &=& \frac{\mu^2}{\left|\bm{r}_{ij}\right|^3} \left( s_i s_j c - 2 c_i
c_j \right), \0\0\0\0\0\0\0\0\0\0\0\0\0\0\0\0\0\0\0\0\0\0\0\0\0\0\0 \label{uD}
\end{eqnarray}
secondly, the interaction between quadrupoles
\begin{eqnarray}
u_{\rm QQ}(\bm{r}_{ij},\bm{\omega}_i,\bm{\omega}_j,Q)   &=& \frac{3}{4} \frac{Q^2}{\left|\bm{r}_{ij}\right|^5} \left[ 1-5
\left( c_i^2+c_j^2 \right) -15 c_i^2 c_j^2 + 2 \left( s_i s_j c - 4 c_i c_j \right)^2 \right], \label{uQ}
\end{eqnarray}
and finally, the mutual interaction between dipoles and quadrupoles 
\begin{eqnarray}
u_{\rm DQ}(\bm{r}_{ij},\bm{\omega}_i,\bm{\omega}_j,\mu,Q)&=& \frac{3}{2} \frac{\mu Q}{\left|\bm{r}_{ij}\right|^4} \left(
c_i-c_j \right)\left( 1 + 3c_i c_j - 2s_i s_j c \right), \0\0\0\0\0\0\0\0\0\0\0\0\0\0\0\0\0 \label{uDQ}
\end{eqnarray}
where $c_k={\rm cos} \theta_k$, $s_k={\rm sin} \theta_k$, and $c={\rm cos} \phi_{ij}$. $\theta_i$ is the angle between the axis of the molecule $i$ and
the center-center connection line and $\phi_{ij}$ is the azimuthal angle between the axis of molecule $i$ and $j$. The
Lennard-Jones parameters $\sigma$ and $\epsilon$ represent size and energy, respectively.
 
Among the different spatial arrangements of the four charges in a quadrupole, in 2CLJDQ models they are arranged along the
molecular axis in the symmetric sequence $+$, $-$, $-$, $+$ or, having the same energetic effect in pure quadrupolar fluids,
$-$, $+$, $+$, $-$. The point quadrupole interaction, cf.~Eq.~(\ref{uQ}), is a large distance approximation, 
reducing the number of parameters related to the quadrupole to one, namely the quadrupolar moment $Q$. The dipole is approximated in the same sense.

In case of elongated fluids, for very small intermolecular distances $|\bm{r}_{ij}|$, the positive Len\-nard-Jones part $u_{\rm 2CLJ}$ of the full potential
cannot outweigh the divergence to $-\infty$ of the polar part $u_{\rm DD}$+$u_{\rm QQ}$+$u_{\rm DQ}$. This divergence of $u_{\rm 2CLJDQ}$ leads to infinite Boltzmann factors, i.e. non-existence of the configurational integral. During molecular dynamics phase space sampling within the pressure range in question, this artifact of the 2CLJDQ potential causes no problem as intermolecular center-center distances are very improbable to fall below critical values.
However, during Monte-Carlo simulation or the calculation of entropic properties by test particle
insertion \cite{widom6328}, critical intermolecular center-center distances might occur.  
To avoid this problem, following M\"oller and Fischer \cite{moeller9435}, a hard sphere of diameter $0.4 \sigma$ was placed directly on the polar sites as a shield for critical configurations.

The parameters $\sigma$ and $\epsilon$ were used for the reduction of all thermodynamic properties as well as the model parameters: $T^*=Tk_B/\epsilon$, $p^*=p\sigma^3/\epsilon$,
$\rho^*=\rho\sigma^3$, $h^*=h/\epsilon$, $L^*=L/\sigma$, $\mu^{*2}=\mu^2/\left(\epsilon \sigma^3 \right)$ and $Q^{*2}=Q^2/\left(\epsilon\sigma^5 \right)$.
The model parameters were varied in this investigation: $L^*=0$, $0.505$ and $1$, $\mu^{*2}=6$ and $12$ as well as $Q^{*2}=2$ and $4$. Combining these values leads to a set of twelve model fluids that were investigated here.

To achieve a monotonous transition from $L^*>0$ to $L^*=0$, spherical fluids were treated as 2CLJ fluids with $L^*=0$. 
This leads to a site superposition that is not present in the one-center Lennard-Jones case. 
Therefore, in reduced units, temperature, pressure and enthalpy are fourfold and the squared polar moments twofold of the corresponding values when only one Lennard-Jones site is present. Densities are not affected.

\section{Molecular simulation method}
{\label{MSMFVD}}
\baselineskip24pt
Throughout all pure fluid VLE simulations, the $N\!pT$ + test particle method by M\"oller and Fischer \cite{moeller9046, moeller9214} was used. 
The chemical potential was calculated in the $N\!pT$ ensemble by Widom's method \cite{widom6328}.

In all simulations a center-center cut-off radius $r_{\rm c}=5\sigma$ was used. 
Outside the cut-off sphere, the fluid was assumed to have no preferential relative orientation of the molecules, i.e. for the calculation of the Lennard-Jones long range corrections, orientational averaging was done with equally weighted relative orientations as proposed by Lustig \cite{lustig8817}. 
Long range corrections for the dipolar interaction were calculated with the reaction field method \cite{barker7378,saager9127}, where the relative permittivity $\epsilon_{\rm s}$ was set to infinity. 
The quadrupolar interaction needs no long range correction as it disappears by orientational averaging. The same holds for the mixed polar interaction between dipoles and quadrupoles, 
cf.~Weingerl and Fischer \cite{weingerl202}.

Configuration space sampling was done with $N=864$ particles for both liquid and vapor simulations in the $NpT$ ensemble 
with Andersons barostat \cite{andersen}.
The reduced integration time step was set to $\Delta t\sqrt{m/\epsilon}/\sigma=0.0015$ and
the reduced membrane mass parameter of the barostat was set to $2 \cdot 10^{-4}$ for liquid and to $10^{-6}$ for vapor
simulations.   
 
Starting from a face centered lattice arrangement, every run was equilibrated over $10,000$ time steps. 
Data production was performed over $100,000$ time steps. For vapor simulations with $L^*=1$, especially those at low temperatures,
the Monte-Carlo method was used to achieve equilibrium within an acceptable time range. 
The number of Monte-Carlo loops was generally chosen to be the same as the number of MD time steps. 
One Monte Carlo loop is defined here as $N$ trial translations, $\left(2/3\right)N$ trial rotations, and one trial volume change.
At each production time step $2N$ test particles were inserted in the liquid phase, and $N$ test particles in the vapor phase
to calculate the chemical potential. To get a better accuracy of the chemical potential, $4N$ test particles were used in
simulations in the liquid phase at low temperatures. 

In some cases, where a highly dense and strongly polar liquid phase was present, the
more elaborate gradual insertion scheme had to be employed to obtain the chemical potential sufficiently
accurate.   
The gradual insertion method is an expanded ensemble method \cite{shevkunov8824} based on the Monte Carlo technique. 
Here, the version proposed by Nezbeda and Kolafa \cite{nezbeda9139} was used, in a form that was extended to the $N\!pT$ ensemble
\cite{vrabec0243}.
In comparison to
Widom's test particle method, where whole molecules are inserted in the fluid, gradual insertion introduces one fluctuating
molecule,
that undergoes changes in a predefined set of discrete states of coupling with the other molecules of the fluid. Preferential sampling
is done in the vicinity of the fluctuating particle. This concept leads to considerably improved accuracy of the residual chemical
potential. Gradual insertion simulations were performed with $N=864$ particles in the liquid phase.
Starting from a face-centered lattice
arrangement every simulation run was given $5,000$ Monte Carlo loops to equilibrate. Data production was performed over $100,000$ Monte
Carlo loops. Further simulation parameters for runs with gradual insertion were taken from Vrabec et al. \cite{vrabec0243}. 

VLE data were determined for temperatures of about 55 \% to 95 \% of $T^*_{\rm c}(\mu^{*2},Q^{*2},L^*)$.
In the whole temperature range all thermodynamic properties of both phases were obtained by simulation.

\section{Simulation results of VLE data}
\baselineskip24pt
Table \ref{t2cljdqvle} reports the VLE data of the regarded twelve model fluids. Vapor pressure $p_{\sigma}^*$, saturated liquid
density $\rho'^*$, saturated vapor density $\rho''^*$, residual saturated liquid enthalpy $h'^{\mbox{\scriptsize res*}}$ and 
residual saturated vapor enthalpy $h''^{\mbox{\scriptsize res*}}$ are presented, where $h^{\mbox{\scriptsize res*}}=h^{\mbox{\scriptsize *}}(T,p)- h^{\mbox{\scriptsize id}}(T)$. 
Statistical uncertainties were determined with the method of Fincham et al. \cite{fincham8645} and the error propagation law.

\begin{figure}[ht]
\begin{center}
\epsfig{file=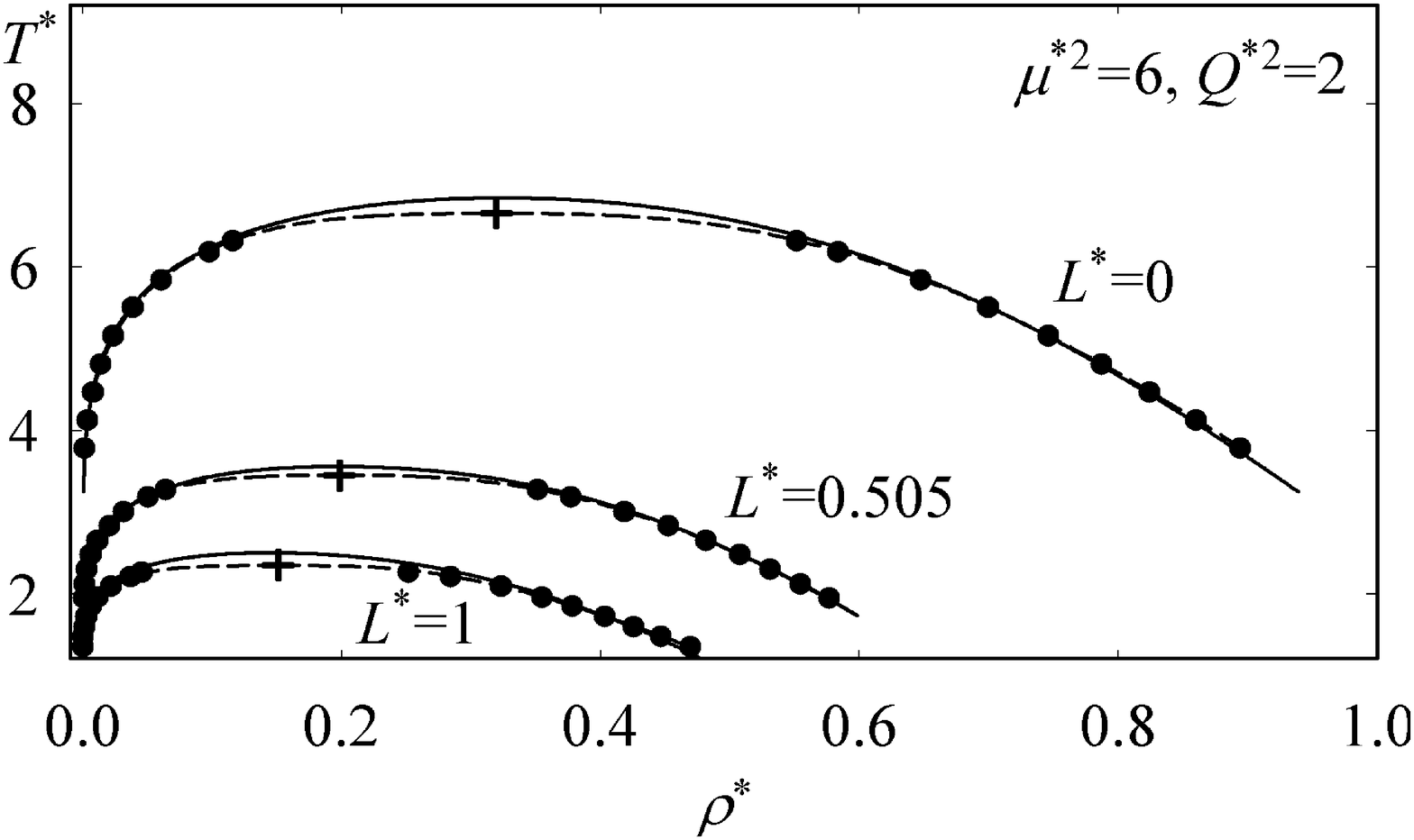, scale=0.47} 
\end{center}
\vskip1cm
\caption[Saturated densities of three 2CLJDQ model fluids with $\mu^{*2}=6$, $Q^{*2}=2$:
{\large $\bullet$}~simulation data, {\bf -----}~present EOS, 
{\bf - - -}~fits to simulation data, cf.~Eqs.~(\ref{Trho1corrcore}) and (\ref{Trho2corrcore}), $\bm +$~critical
points.]{Vrabec and Gross} \label{Tr62}
\end{figure}

\begin{figure}[ht]
\begin{center}
\epsfig{file=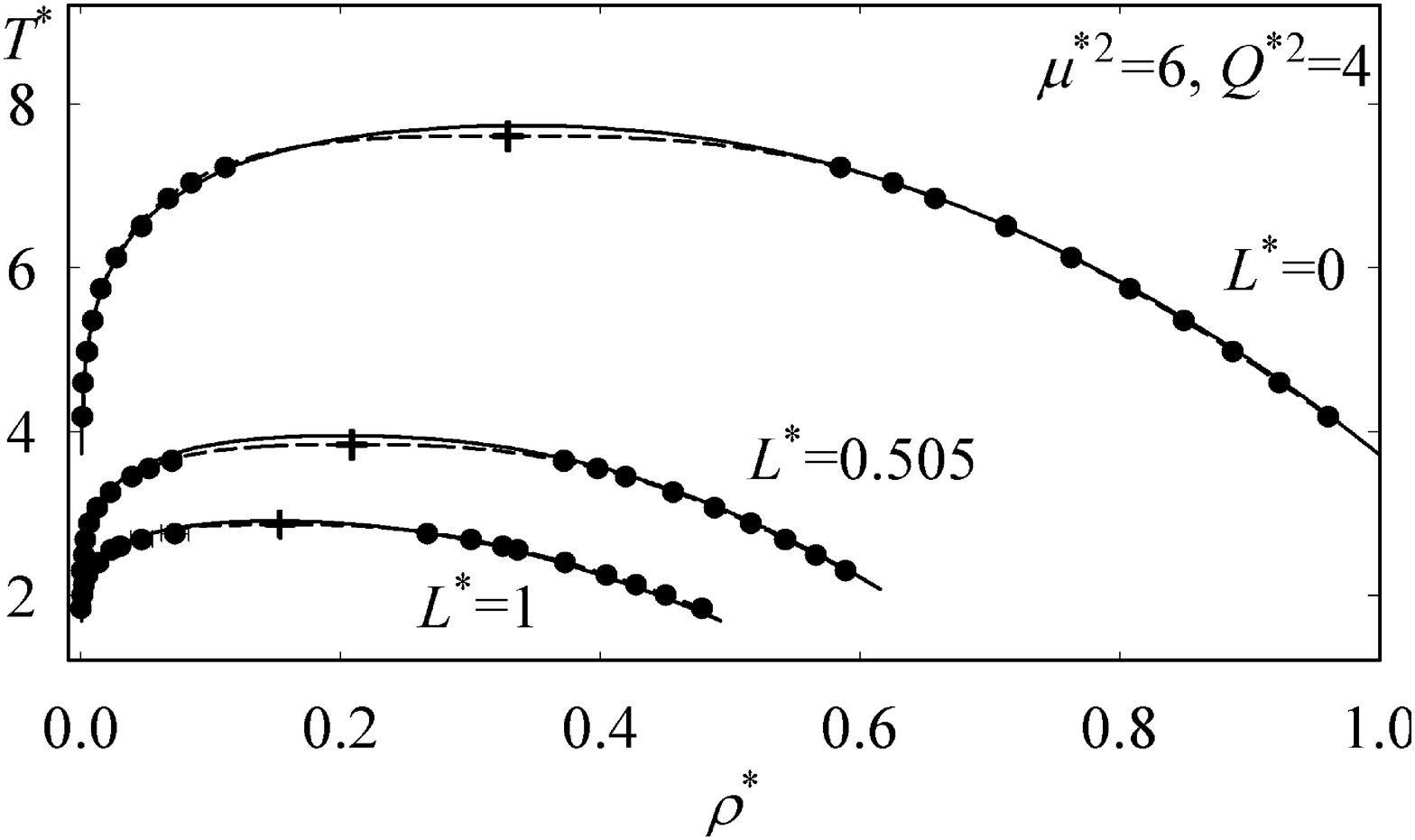, scale=0.47} 
\end{center}
\vskip1cm
\caption[Saturated densities of three 2CLJDQ model fluids with $\mu^{*2}=6$, $Q^{*2}=4$:
{\large $\bullet$}~simulation data, {\bf -----}~present EOS, 
{\bf - - -}~fits to simulation data, cf.~Eqs.~(\ref{Trho1corrcore}) and (\ref{Trho2corrcore}), $\bm +$~critical
points.]{Vrabec and Gross} \label{Tr64}
\end{figure}

\begin{figure}[ht]
\begin{center}
\epsfig{file=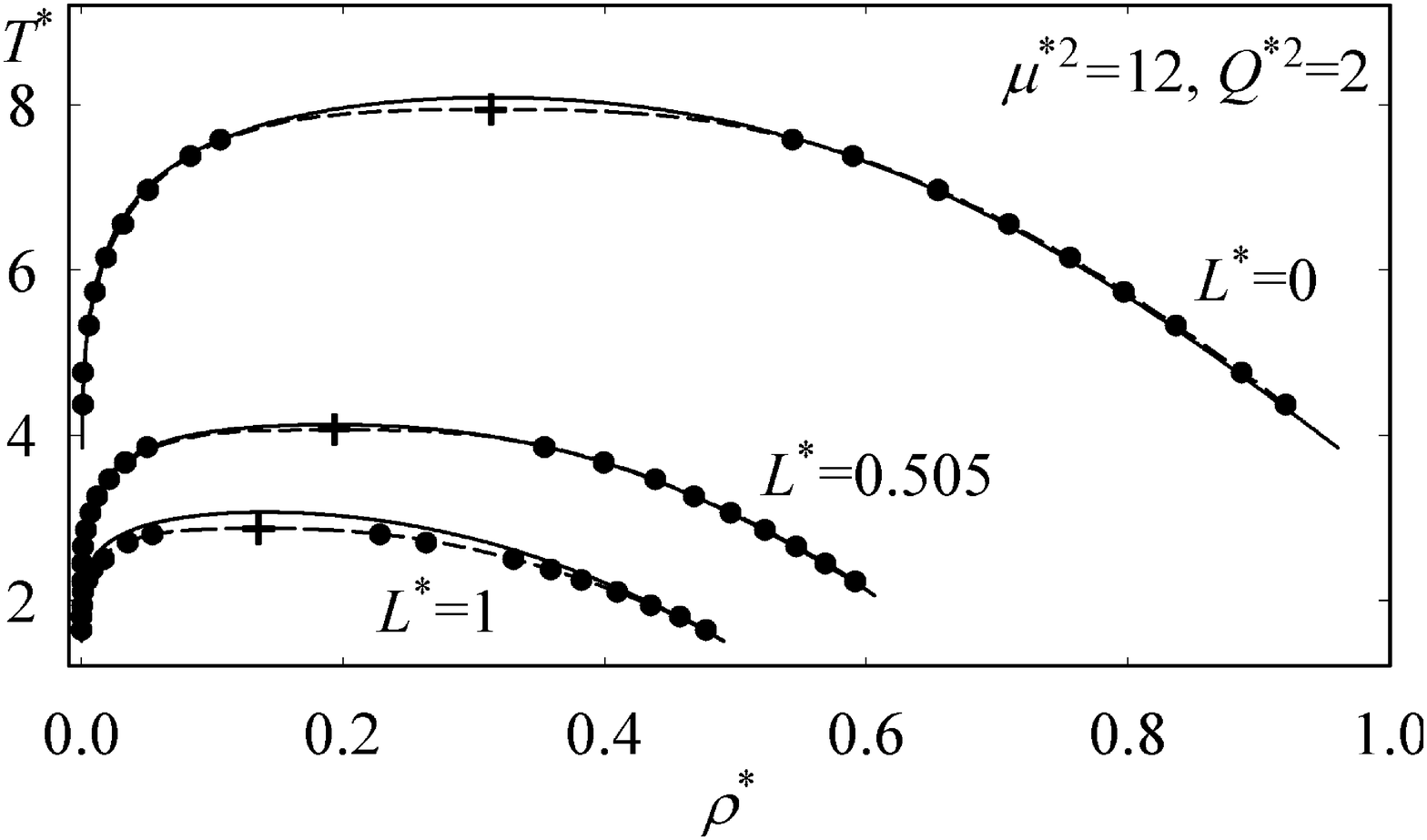, scale=0.47} 
\end{center}
\vskip1cm
\caption[Saturated densities of three 2CLJDQ model fluids with $\mu^{*2}=12$, $Q^{*2}=2$:
{\large $\bullet$}~simulation data, {\bf -----}~present EOS, 
{\bf - - -}~fits to simulation data, cf.~Eqs.~(\ref{Trho1corrcore}) and (\ref{Trho2corrcore}), $\bm +$~critical
points.] {Vrabec and Gross} \label{Tr122}
\end{figure}

\begin{figure}[ht]
\begin{center}
\epsfig{file=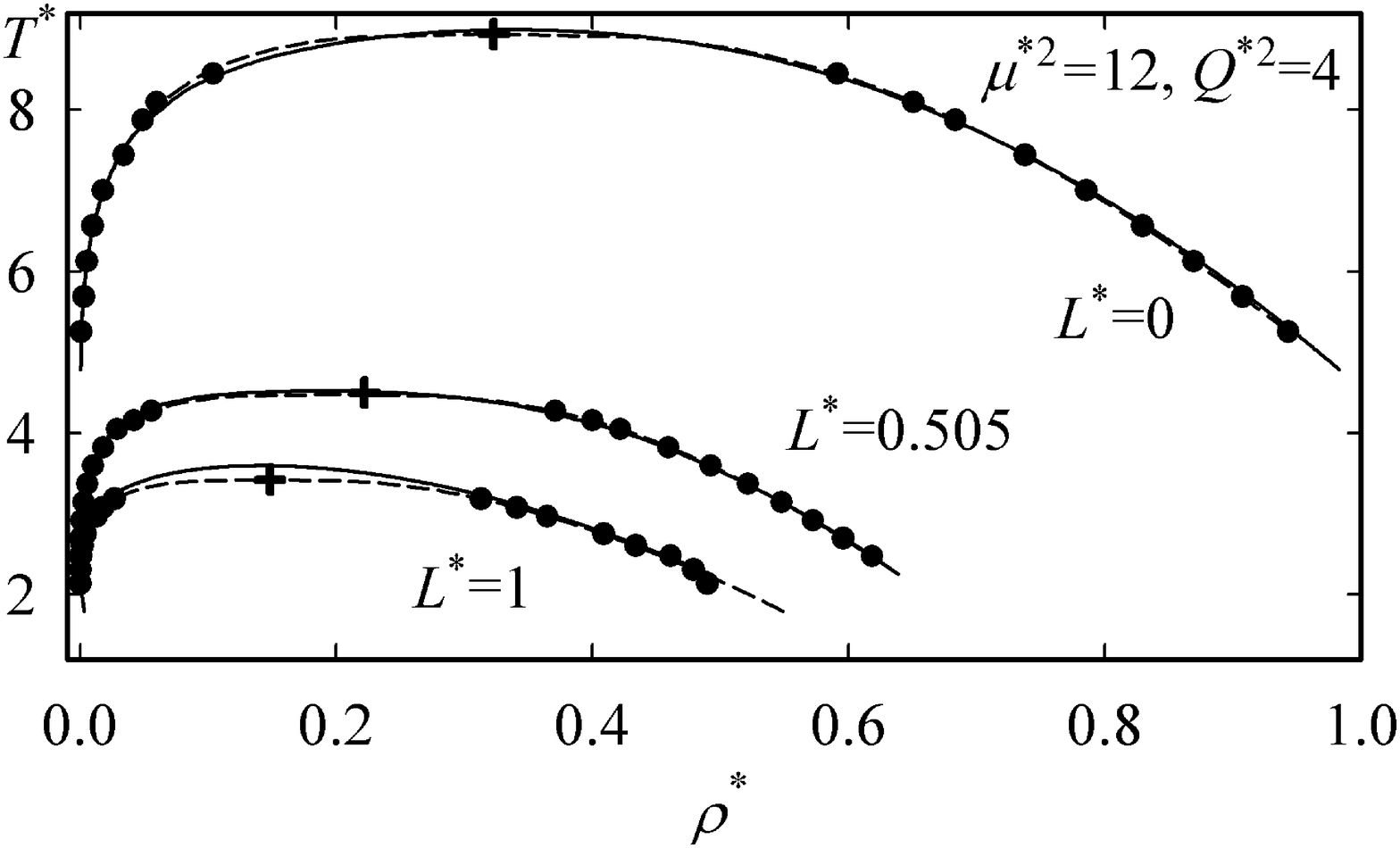, scale=0.47} 
\end{center}
\vskip1cm
\caption[Saturated densities of three 2CLJDQ model fluids with $\mu^{*2}=12$, $Q^{*2}=4$:
{\large $\bullet$}~simulation data, {\bf -----}~present EOS, 
{\bf - - -}~fits to simulation data, cf.~Eqs.~(\ref{Trho1corrcore}) and (\ref{Trho2corrcore}), $\bm +$~critical
points.]{Vrabec and Gross} \label{Tr124}
\end{figure}

\begin{figure}[ht]
\begin{center}
\epsfig{file=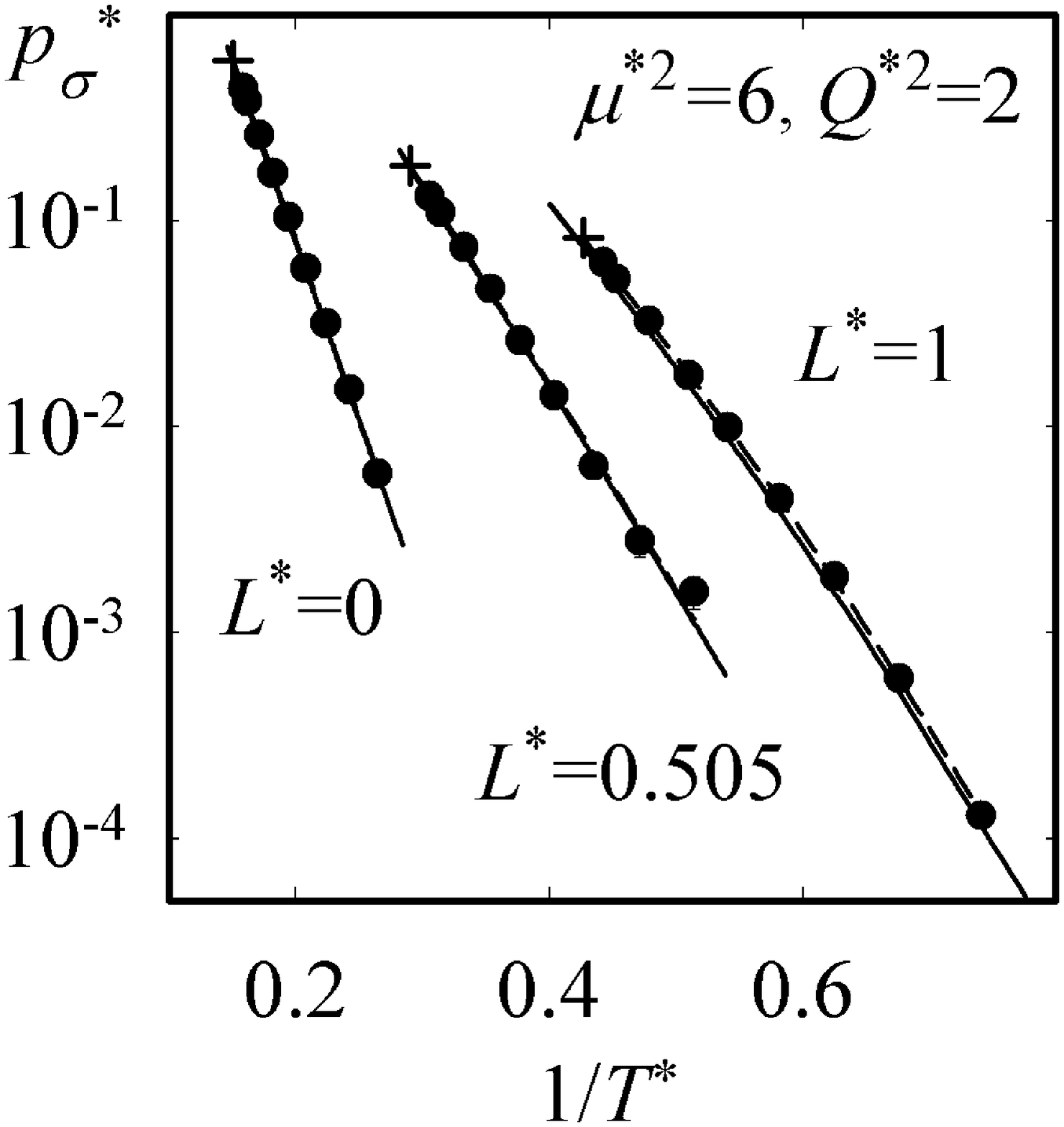, scale=0.47} 
\end{center}
\vskip1cm
\caption[Vapor pressure of three 2CLJDQ model fluids with $\mu^{*2}=6$, $Q^{*2}=2$:
{\large $\bullet$}~simulation data, {\bf -----}~present EOS, 
{\bf - - -}~fits to simulation data, cf.~Eq.~(\ref{pTcorr}), $\bm +$~critical
points.]{Vrabec and Gross} \label{Tp62}
\end{figure}

\begin{figure}[ht]
\begin{center}
\epsfig{file=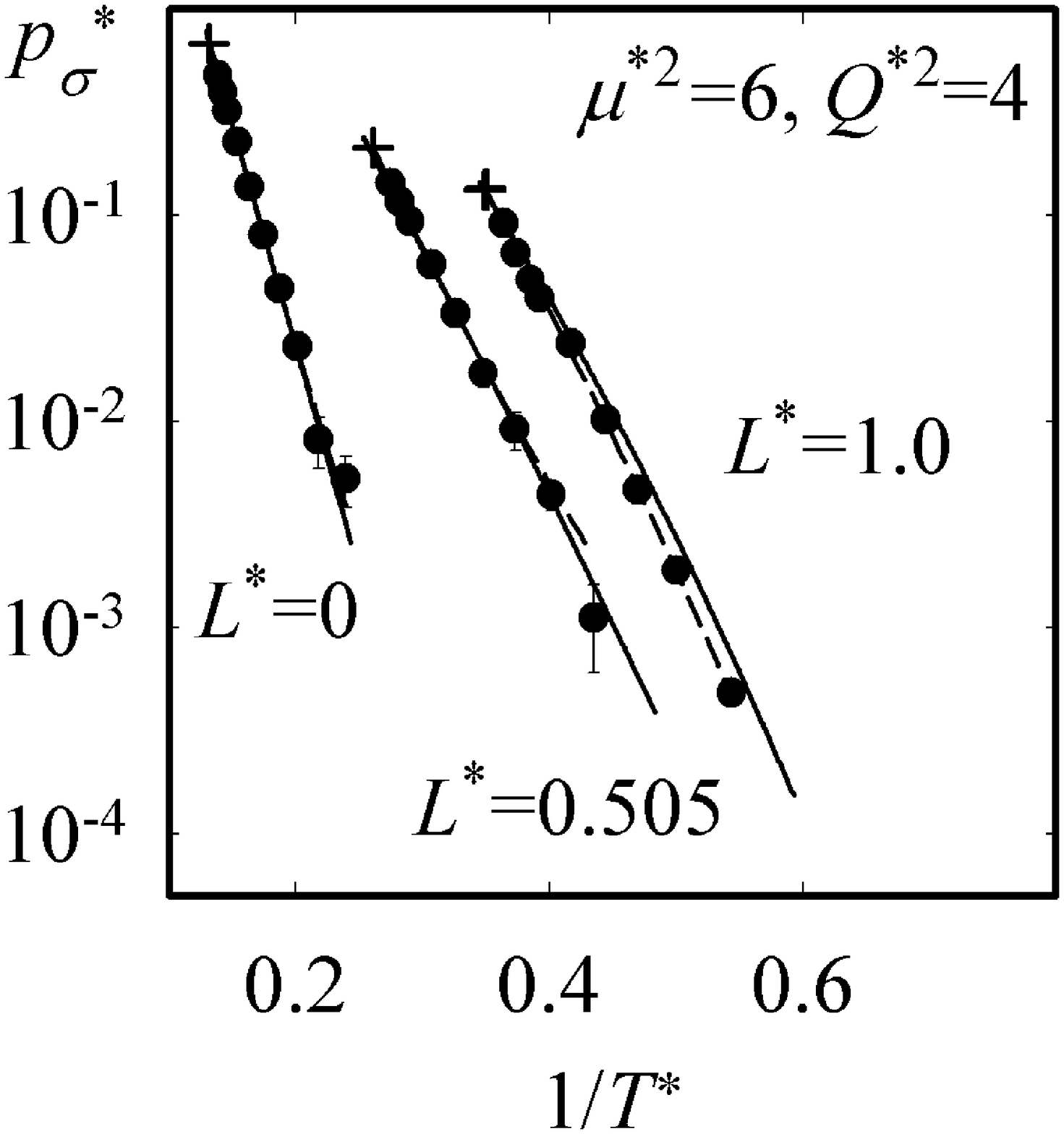, scale=0.47} 
\end{center}
\vskip1cm
\caption[Vapor pressure of three 2CLJDQ model fluids with $\mu^{*2}=6$, $Q^{*2}=4$:
{\large $\bullet$}~simulation data, {\bf -----}~present EOS, 
{\bf - - -}~fits to simulation data, cf.~Eq.~(\ref{pTcorr}), $\bm +$~critical
points.]{Vrabec and Gross} \label{Tp64}
\end{figure}

\begin{figure}[ht]
\begin{center}
\epsfig{file=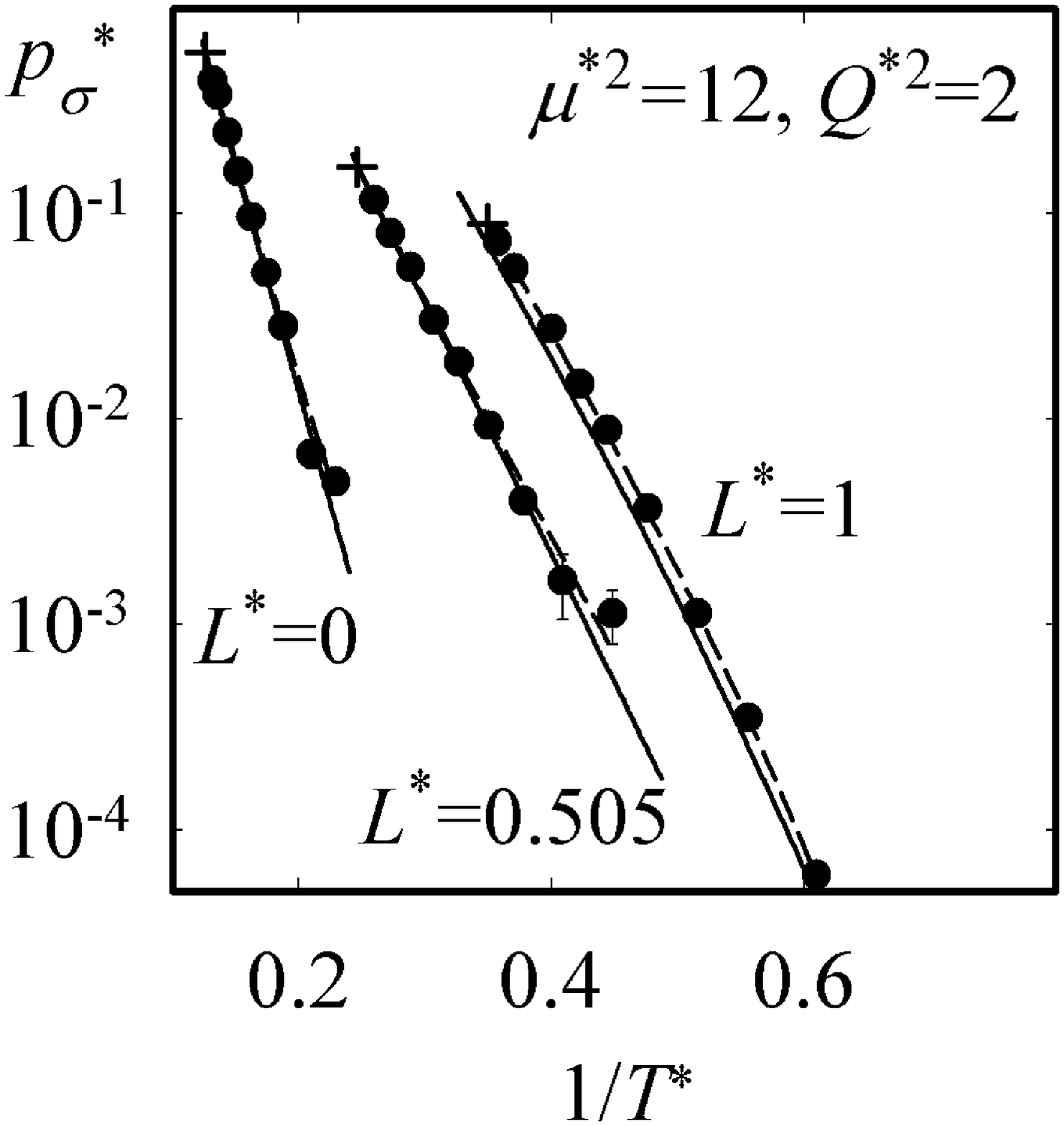, scale=0.47}
\end{center}
\vskip1cm
\caption[Vapor pressure of three 2CLJDQ model fluids with $\mu^{*2}=12$, $Q^{*2}=2$:
{\large $\bullet$}~simulation data, {\bf -----}~present EOS, 
{\bf - - -}~fits to simulation data, cf.~Eq.~(\ref{pTcorr}), $\bm +$~critical
points.]{Vrabec and Gross} \label{Tp122}
\end{figure}

\begin{figure}[ht]
\begin{center}
\epsfig{file=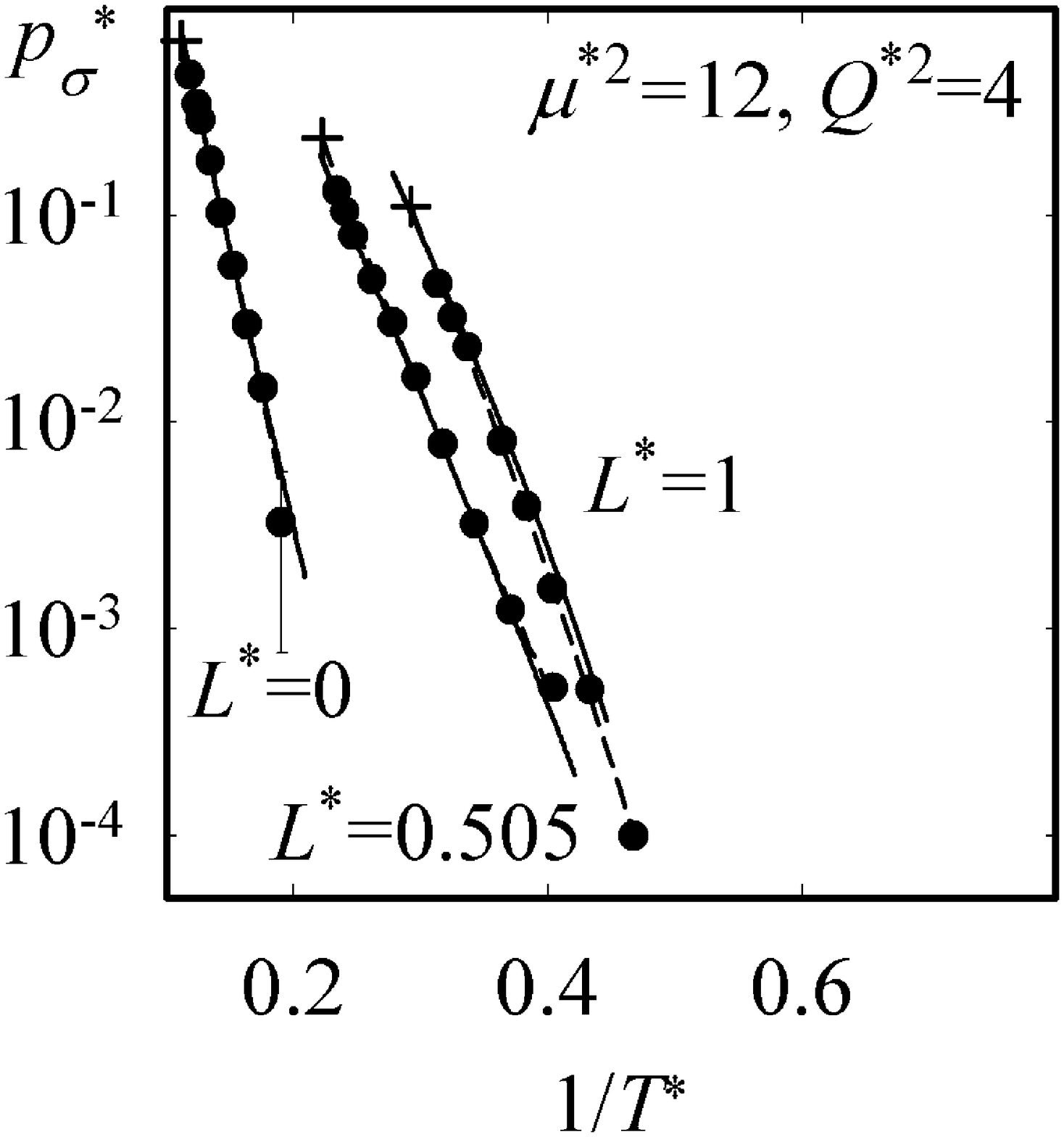, scale=0.47}
\end{center}
\vskip1cm
\caption[Vapor pressure of three 2CLJDQ model fluids with $\mu^{*2}=12$, $Q^{*2}=4$:
{\large $\bullet$}~simulation data, {\bf -----}~present EOS, 
{\bf - - -}~fits to simulation data, cf.~Eq.~(\ref{pTcorr}), $\bm +$~critical
points.]{Vrabec and Gross} \label{Tp124}
\end{figure}

Figs.~\ref{Tr62} to \ref{Tp124} illustrate the strong influence of both 
elongation and polar moments on the VLE data. Elongation and polar moments strongly influence saturated density and the vapor pressure curves. 
As can be expected, all three critical properties in Figs.~\ref{Tr62} to \ref{Tp124}, i.e. temperature, density and pressure, decrease with increasing elongation 
for a given set of polar moments and thus the phase envelopes migrate accordingly. 
For a constant elongation, they usually increase with increasing polar moments. 

To determine the critical data quantitatively, the method of Lotfi et al. \cite{lotfi9213} was used.
The saturated density--temperature dependence near the critical point is well described by $\rho^* \sim \left( T^*_{\rm c}-T^*\right)^{1/3}$, as given by Guggenheim \cite{guggenheim4525,
rowlinson1969}. Therefore, the following correlations were fitted to simulated vapor pressure and saturated densities
\begin{eqnarray}
\ln p^*_{\sigma} &=& c_1 + \frac{c_2}{T^*}+\frac{c_3}{T^{*4}}, \label{pTcorr}\\
\rho'^* &=& \rho^*_{\rm c} + C_1 \cdot (T^*_{\rm c}-T^* )^{1/3} + C'_2 \cdot (T^*_{\rm c}-T^*) + C'_3 \cdot
(T^*_{\rm c}-T^*)^{3/2},
\label{Trho1corrcore} \\
\rho''^* &=& \rho^*_{\rm c} - C_1 \cdot (T^*_{\rm c}-T^* )^{1/3} + C''_2 \cdot (T^*_{\rm c}-T^*) + C''_3 \cdot (T^*_{\rm
c}-T^*)^{3/2}. \label{Trho2corrcore}
\end{eqnarray}
The simultaneous fit of saturated densities, cf.~Eqs.~(\ref{Trho1corrcore}) and (\ref{Trho2corrcore}), yields $\rho^*_{\rm c}$ and $T^*_{\rm
c}$. Inserting $T^*_{\rm c}$ into the vapor pressure fit, cf.~Eq.~(\ref{pTcorr}), gives the critical pressure $p^*_{\rm c}$. These fits can be seen in Figs.~\ref{Tr62} to \ref{Tp124}.
The critical data for the regarded twelve 2CLJDQ model fluids are given numerically in Table \ref{2CLJDQcdomtab}, which also contains the acentric factor $\omega$
\begin{eqnarray}
\omega \left( \mu^{*2},Q^{*2},L^* \right)=-{\rm log}_{10}\frac{p_{\sigma}^* \left( \mu^{*2},Q^{*2},L^*,0.7T^*_{\rm
c}\right)}{p^*_{\rm c} \left( \mu^{*2},Q^{*2},L^* \right)}-1, \label{omega}
\end{eqnarray}
and the critical compressibility $Z_{\rm c}=p^*_{\rm c}/\left(\rho^*_{\rm c} T^*_{\rm c}\right)$. 

\begin{figure}[ht]
\begin{center}
\epsfig{file=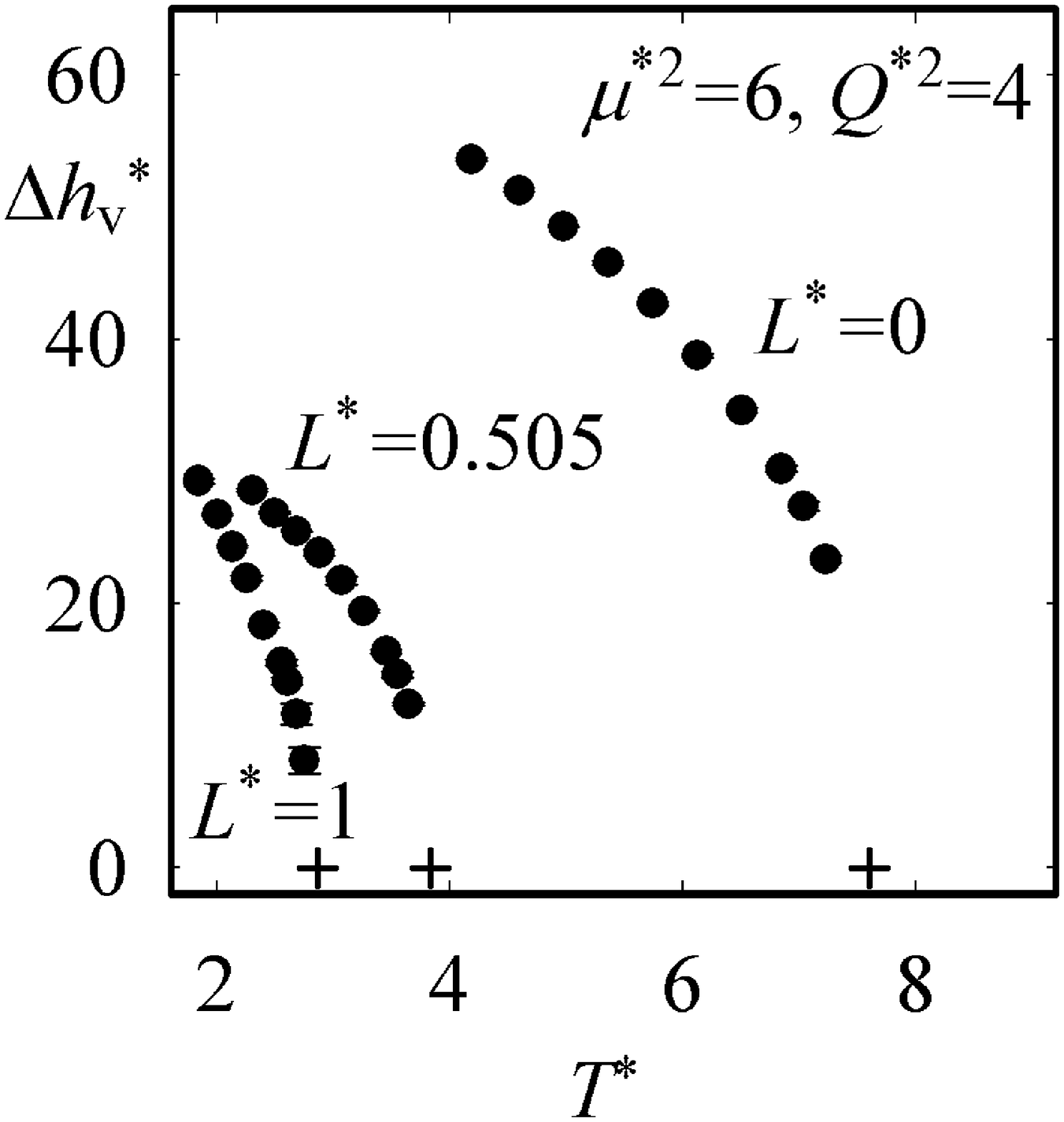, scale=0.47} 
\end{center}
\vskip1cm
\caption[Enthalpy of vaporization of three 2CLJDQ model fluids with $\mu^{*2}=6$, $Q^{*2}=4$:
{\large $\bullet$}~simulation data, $\bm +$~critical points.] {Vrabec and Gross} \label{dhv64}
\end{figure}

Fig.~\ref{dhv64} illustrates the influence of the elongation on the enthalpy of vaporization. An increasing elongation reduces the enthalpy of vaporization. As can be expected
(and not shown here), increasing polar moments increase the enthalpy of vaporization.  

Thermodynamic consistency of the simulation data was validated with the Clausius-Clapeyron equation\\
\begin{eqnarray}
\frac{\partial {\rm ln}p_{\sigma}}{\partial T}=\frac{\Delta h_{\rm v}}{p_{\sigma} T \left( 1/\rho''-1/\rho'
\right)}. \label{clcl}
\end{eqnarray}\\
The vapor pressure correlation, cf.~Eq.~(\ref{pTcorr}), was used for the left hand side of Eq.~(\ref{clcl}), while the right hand side was calculated from present simulation data. 
Within statistical uncertainties, Eq.~(\ref{clcl}) is fulfilled almost throughout.

\begin{figure}[ht]
\begin{center}
\epsfig{file=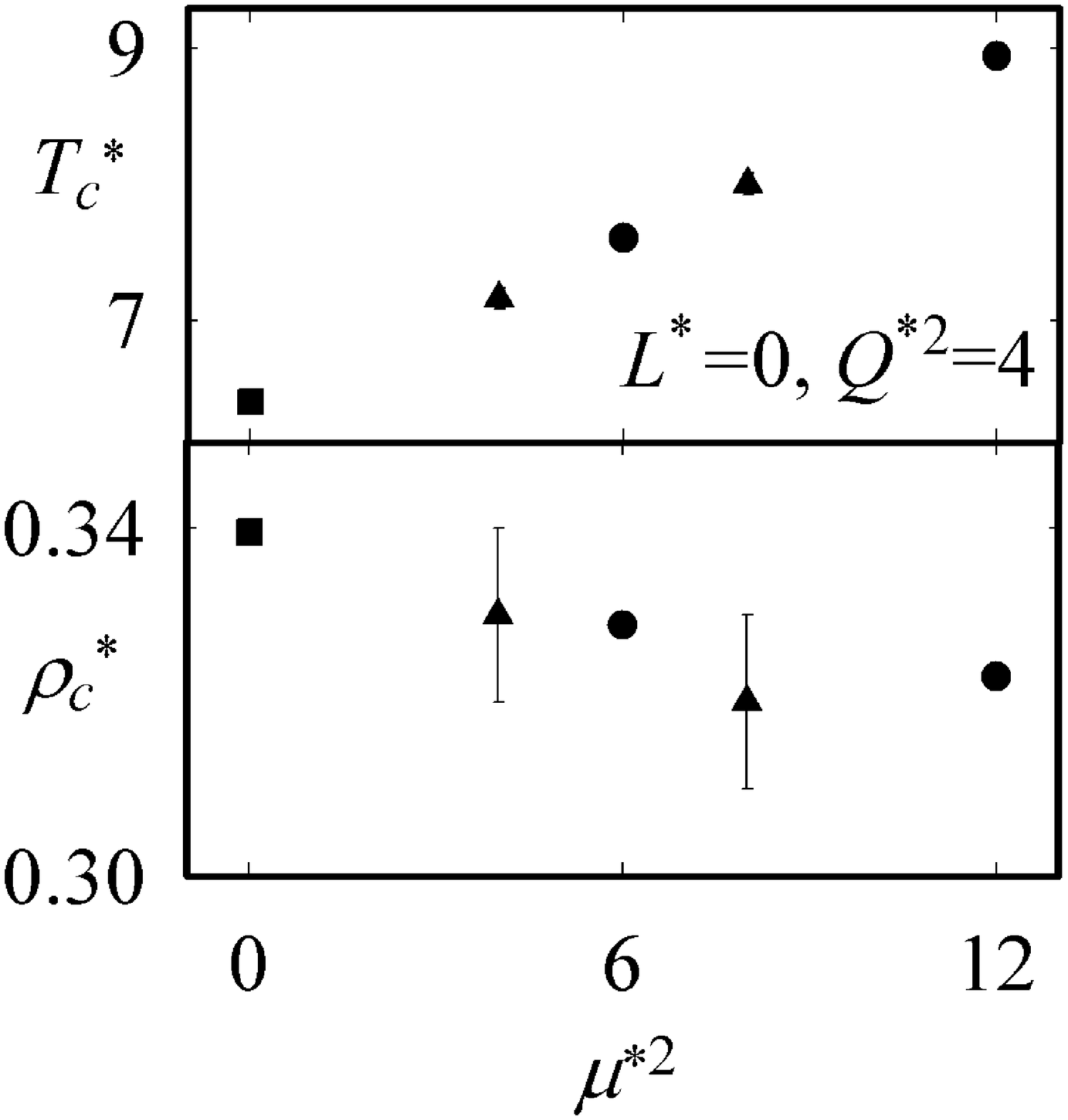, scale=0.47}
\end{center}
\vskip1cm
\caption[Critical temperatures (top) and densities (bottom) for five 2CLJDQ model fluids with
$L^*=0$, $Q^{*2}=4$ from simulation:
{\large $\bullet$}~present data, $\blacktriangle$~Dubey and O'Shea 
\cite{dubey9421}, {\Huge $\centerdot$}~Stoll et al. \cite{stoll4549}.]{Vrabec and Gross}
\label{crit}
\end{figure}

Results from the present study can be compared with other data. 
Dubey and O'Shea \cite{dubey9421} studied VLE of model fluids with both point dipoles and point quadrupoles embedded in one Lennard-Jones site, but unfortunately out of our grid of
molecular parameters. 
In previous work of our group \cite{stoll4549}, VLE of 2CLJQ model fluids were studied, which is a limiting case here. 
Fig.~\ref{crit} presents a comparison of the critical temperature and critical density for different $\mu^{*2}$ for a specified combination of $L^*=0$ and $Q^{*2}=4$. 
It can be seen that present 2CLJDQ data agrees well with the other critical data 
(including Dubey and O'Shea) \cite{stoll4549,dubey9421}.

\section{Equation of state theory}
{\label{EOST}}
\baselineskip24pt

An EOS for 2CLJDQ fluids can be developed with the help of the VLE simulation results 
discussed above by applying a perturbation theory following the intermolecular
potential, cf. Eqs.~(\ref{fullpotential}) and (\ref{Ares}). Using the non-polar 2CLJ fluid as a reference fluid,
each of the polar contributions
to the intermolecular potential, i.e. $u^{DD}$, $u^{QQ}$ and $u^{DQ}$, leads to a perturbation expansion
$A^{DD}$, $A^{QQ}$ and $A^{DQ}$ of the Helmholtz energy $A$. The EOS written in the residual Helmholtz energy then reads
\begin{equation}
\frac{A^{res}}{Nk_BT} = \frac{A^{2CLJ}}{Nk_BT} + \frac{A^{DD}}{Nk_BT} + \frac{A^{QQ}}{Nk_BT} + \frac{A^{DQ}}{Nk_BT}, \0 \label{Ares}
\end{equation}
where $A^{2CLJ}$ is the residual Helmholtz energy of the 2CLJ reference fluid.
The EOS of this reference fluid, $A^{2CLJ}$, can easily be constructed with Wertheim's Thermodynamic 
Perturbation Theory of first order (TPT1) \cite{wertheim8419,wertheim8435,wertheim86459,wertheim86477}.
Although the TPT1 initially
resulted in a description of tangent-sphere fluids, rather than fused-sphere configurations 
(such as the 2CLJ model), the TPT1 was shown to form a framework, where the 2CLJ
fluid behavior can be recovered with high accuracy. A relation mapping the elongation 
$L^*$ of the 2CLJ fluid to the (noninteger) "{}number of tangent-spheres"{} $m$ was earlier reported \cite{gross05}.
The other parameters of 2CLJDQ fluids, $\sigma$, $\epsilon$, $\mu^*$ and $Q^*$, also need to be converted to 
those of the tangent-sphere model. This conversion is straightforward and it is referred to Table 1
of ref. \cite{gross05}. The superscript "{}TS"{} is used  here to indicate parameters of the 
tangent-sphere model, e.g. it is $\sigma=\sigma^{TS}$. The EOS of the 2CLJ fluid is then 
\begin{equation}
\frac{A^{2CLJ}}{Nk_BT} = m \cdot \frac{A^{LJ}}{Nk_BT} + (1 - m) \cdot \ln g^{LJ}(\sigma),
\end{equation}
where $g^{LJ}(\sigma)$ is the value of the radial distribution function at the distance $r = \sigma$ as proposed by Johnson et
al. \cite{johnson94}.

Expressions for $A^{QQ}$ and $A^{DD}$ where reported earlier \cite{gross05,gross06} as third order perturbation
expansions written in the Pad\'e approximation. To develop an expression for $A^{DQ}$, an analogous
procedure is followed here. The EOS contribution $A^{DQ}$ is presented below for the case of
mixtures, since mixtures are considered in section 6.
The Helmholtz energy contribution $A^{DQ}$ to Eq.~(\ref{Ares}) is given as 
\begin{equation}
\frac{A^{DQ}}{Nk_BT} = \frac{{A_{2}^{DQ}}/{Nk_BT}}{1 - {A_{3}^{DQ}}/{A_{2}^{DQ}}}, \0 \label{Adq}
\end{equation}
with $A_{2}^{DQ}$ and $A_{3}^{DQ}$ being the second and third order perturbation terms, respectively. 
For linear and symmetric molecules, which are considered here, the second order term $A_{2}^{DQ}$ reads in dimensionless form
\begin{equation}
\frac{A_{2}^{DQ}}{Nk_BT} = - \frac{9\pi}{4} \rho \sum_{i} \sum_{j} x_{i} x_{j} \frac{\epsilon_{ii}^{TS}}{k_BT}
\frac{\epsilon_{jj}^{TS}}{k_BT} \frac{\sigma_{ii}^{3} \sigma_{jj}^{5}}{\sigma_{ij}^{5}} \mu_{i}^{*TS2} Q_{j}^{*TS2}
J_{2,ij}^{DQ}, \0 \label{A2dq} 
\end{equation}
where $x_{i}$ is the mole fraction of component $i$, 
$\mu_i^{*TS2} = \mu_i^2/(m_i \epsilon_{ii}^{TS} \sigma_{ii}^3$) denotes the dimensionless squared dipole moment in the 
tangent sphere framwork, and $Q_i^{*TS2} = Q_i^2/(m_i \epsilon_{ii}^{TS} \sigma_{ii}^5$) the respective dimensionless
squared quadrupole moment.  The unlike Len\-nard-Jones parameters
$\epsilon_{ij}^{TS}$ and $\sigma_{ij}$ were determined according to the Berthelot-Lorentz combing rules, with
$\epsilon_{ij}^{TS} = \left( \epsilon_{i}^{TS} \epsilon_{j}^{TS} \right)^{1/2}$ and $\sigma_{ij} = \left( \sigma_{i} +
\sigma_{j} \right) / 2$. The abbreviation $J_{2,ij}^{DQ}$ denotes integrals over angles and radius with an
integrand being the angle-dependent part of the intermolecular potential weighted with the reference-fluid pair-correlation
function. An analytic expression for $J_{2,ij}^{DQ}$ is not easily available and following
\cite{gross06,gross05}, simple power functions were assumed for $J_{2,ij}^{DQ}$, where the
coefficients were adjusted to the present VLE simulation results. 

The third order term accounts for three-body effects and consists of two contributions $A_{3}^{DQ} = A_{3}^{DDQ}
+ A_3^{DQQ}$. Similar to $J_{2,ij}^{DQ}$ in the second order term, both terms $A_{3}^{DDQ}$ and $A_3^{DQQ}$
contain integrals over the three-body correlation functions. The available data base, including the simulation 
results of this study, is not sufficiently broad to justify an independent adjustment of 
expressions for both of these integrals and therefore the third order term is simplified to 
\begin{eqnarray}
\frac{A_{3}^{DQ}}{Nk_BT} = - \rho^{2} \sum_{i} \sum_{j} \sum_{k} x_{i} x_{j} x_{k}
\frac{\epsilon_{ii}^{TS}}{k_BT} \frac{\epsilon_{jj}^{TS}}{k_BT} \frac{\epsilon_{kk}^{TS}}{k_BT} \frac{\sigma_{ii}^{4} \sigma_{jj}^{4}
\sigma_{kk}^{4}}{\sigma_{ij}^{2} \sigma_{ik}^{2} \sigma_{jk}^{2}} \nonumber \\
\cdot\left( \mu_{i}^{*TS2} \mu_{j}^{*TS2} Q_{k}^{*TS2} +
\alpha \cdot \mu_{i}^{*TS2} Q_{j}^{*TS2} Q_{k}^{*TS2} \right) J_{3,ijk}^{DQ}. \label{A3dq} 
\end{eqnarray}
It is thereby assumed that the integrals $J_{3,ijk}$, which appear in both $A_{3}^{DDQ}$ and
$A_{3}^{DQQ}$, have a similar density dependence and that the difference between them can be captured by an
empirical factor $\alpha$ in Eq.~(\ref{A3dq}). This is certainly a pragmatic approach and may need refinement
once a broader data base of simulation results is available. For the integrals $J_{2,ij}^{DQ}$ and
$J_{3,ijk}^{DQ}$ simple power functions in density and a rudimentary temperature dependence
in the second order term are assumed
\begin{eqnarray}
J_{2,ij}^{DQ} &=& \sum_{n = 0}^{4} \left( a_{n,ij} + b_{n,ij} \frac{\epsilon_{ij}^{TS}}{k_BT} \right) \eta^{n}, \0 \label{J2dd} \\
J_{3,ijk}^{DQ} &=& \sum_{n = 0}^{4} c_{n,ijk} \eta^{n}, \0 \label{J3dd}
\end{eqnarray}
where $\eta$ denotes the packing fraction, which is a dimensionless density. The relation between $\eta$ and $\rho$
depends on the considered EOS, cf. ref. \cite{gross05}. The
coefficients in Eqs.~(\ref{J2dd}) and ~(\ref{J3dd}) depend on the chain length $m$ with 
\begin{eqnarray}
a_{n,ij} &=& a_{0n} + \frac{m_{ij} - 1}{m_{ij}} a_{1n} + \frac{m_{ij} - 1}{m_{ij}} \frac{m_{ij} - 2}{m_{ij}} a_{2n}, \0 \label{anij} \\
b_{n,ij} &=& b_{0n} + \frac{m_{ij} - 1}{m_{ij}} b_{1n} + \frac{m_{ij} - 1}{m_{ij}} \frac{m_{ij} - 2}{m_{ij}} b_{2n}, \0 \label{bnij} \\
c_{n,ijk} &=& c_{0n} + \frac{m_{ijk} - 1}{m_{ijk}} c_{1n}, \0 \label{cnijk}
\end{eqnarray}
and the combining rules of the chain length are given by
\begin{eqnarray}
m_{ij} &=& \left( m_{i} \cdot m_{j} \right)^{1/2} \0 \label{mij} \\
m_{ijk} &=& \left( m_{i} \cdot m_{j} \cdot m_{k} \right)^{1/3} \0 \label{mijk} 
\end{eqnarray}
The simulation data of vapor pressure and saturated densities given in Table \ref{t2cljdqvle} was used to adjust the EOS
constants in Eqs.~(\ref{A3dq}) and (\ref{anij}) to (\ref{cnijk}). The contributions due to the quadrupole-quadrupole
interactions and the dipole-dipole interactions needed for optimizing these constants were taken from our
earlier studies \cite{gross06,gross05}. Eleven constants, namely $\alpha$ in
Eq.~(\ref{A3dq}) as well as
$a_{0n}$, $b_{0n}$ and $c_{0n}$ in Eqs. (\ref{anij}) to (\ref{cnijk}), were adjusted in a first step to the present 
simulation results for spherical 2CLJDQ fluids, i.e. with $L^*=0$. 
The remaining fifteen constants $a_{1n}$,
$a_{2n}$ and $b_{1n}$, $b_{2n}$ and $c_{1n}$ were subsequently adjusted to present simulation results of
elongated 2CLJDQ fluids, cf. Table~\ref{modelconstants}.

For the limiting case of non-polar 2CLJ fluids the EOS overpredicts the critical point. During the adjustement
of the model constants, the critical point was thus not enforced. A comparison of the EOS to simulation data for
saturated liquid and vapor denities is shown in Fig.~\ref{Tr62} to \ref{Tr124}. Apart from systematic deviations
around the critical point, the EOS describes the data with good accuracy. A comparison of the EOS to simulation
data for vapor pressures is given in Fig.~\ref{Tp62} to \ref{Tp124}. Some deviations become apparant for the
highest elongation of $L^*=1$, while the vapor pressure curves of spherical 2CLJDQ fluids are found to be well described.

\section{Application to polar mixtures}
\baselineskip24pt

The proposed EOS can readily be applied to mixtures of dipolar and quadrupolar fluids. 
To assess the capability of the proposed EOS regarding the mixed polar interaction, it was
compared to simulative and experimental VLE of the following five binaries: 
C$_2$H$_2$+R152a, R142b+R113, R12+CO$_2$, R22+R142b and Propylene+R115. 

Five of these components are strongly quadrupolar C$_2$H$_2$ ($Q=5.1$ D\AA), R113 (13.0 D\AA), Propylene
(5.9 D\AA), R115 (9.2 D\AA) and CO$_2$ (3.8 D\AA) and were modelled by 2CLJQ models in \cite{stoll03,vrabec105}. The
numbers in parentheses indicate the polar moment of the molecular model, respectively. 
The remaining four components R152a ($\mu=2.7$ D), R142b (3.0 D), R22 (2.3 D) and R12 (2.3 D) are
dipolar and were modelled by 2CLJD models in \cite{stoll03}.  
Hence, the first three mixtures mentioned above exhibit the dipole-quadrupole
cross-interaction, whereas the remaining two are dipolar or quadrupolar only and are limiting
cases here. Simulative VLE calculations were performed on the basis of these molecular models in
\cite{huang}, where one state independent binary parameter $\xi$ for the unlike Lennard-Jones
interaction  
\begin{equation}
\epsilon_{ij}=\xi \sqrt{\epsilon_i \epsilon_j}.
\label{xiepsLB}
\end{equation}
was adjusted to one experimental vapor pressure following the procedure proposed in \cite{stoll000}.
Table 4 contains the binary parameters derived in \cite{huang} and Table 5 compiles the binary VLE data 
from simulation in numerical form. As can be seen in Figs. 11 to 15, the
simulative approach agrees excellently to the experimental data which exhibits in part a qualitatively different
phase behavior.

The adopted molecular model (2CLJDQ) is of course simple for the considered real fluids; the most delicate assumptions are: Firstly, the polarizability is not accounted for and vacuum-value dipole moments are used. Secondly, the multipole moments are not represented by (distributed) charges but by point-multipoles. Thirdly, the multipolar moments are assumed to be aligned along the molecular axis. The good agreement of the simulation results to the experimental phase behavior however suggests that important characteristics of the real fluids are captured.

To validate the EOS containing the new dipole-quadrupole contribution, these molecular models were used with exactly the same
parameters without any further adjustment. The results are also shown in Figs. 11 to 15.  
It can be seen that the EOS is in excellent agreement with the simulations, usually within the statistical
uncertainty of the simulation data which is often within symbol size. Fig. 12 also gives a comparison with the EOS suggested by Weingerl and Fischer \cite{weingerl202}. This model (dotted line) is found in good but not entirely quantitative agreement.

The components of, e.g., the mixture R142b+R113 have high dipole and quadrupole moments, leading to considerable dipole-quadrupole cross-interactions in the mixture. The
observation that the EOS predicts the mixture phase behavior well suggests that the polar
cross-interactions are adequately captured by the new EOS model. This conclusion is supported by a calculation also shown in Fig. 12, where
the dipole-quadrupole cross-iteractions between the two compounds are set to zero. This is done in order to illustrate the contribution that is due to the here proposed expression for $A^{DQ}$. The EOS is for that case (dashed line) not in the vicinity of the data.

\section{Conclusion}
\baselineskip24pt

The VLE of dipolar and quadrupolar two-center Lennard-Jones fluids was
studied by molecular simulation to derive a new EOS contribution for the
dipole-quadrupole interactions of non-spherical fluids. The vapor pressure, saturated densities and
saturated enthalpies were determined with the $N\!pT$ + test particle method applying a gradual
insertion scheme in cases of high polar densities. The EOS expression has the form of a perturbation
theory of third order where coefficients were adjusted to the bulk properties of the fluids considered
in this work. The EOS was subsequently applied to mixtures without adjustable parameters where an
excellent agreement to experiment and simulation data was found.

\section{Acknowledgment}
\baselineskip24pt

We gratefully acknowledge financial support by Deutsche Forschungsgemeinschaft, Schwer\-punktprogramm 1155. 
The simulations were performed on the national super computer NEC SX-8 at the High Performance Computing Center Stuttgart (HLRS) under the grant MMHBF.

\clearpage

\section*{List of Symbols}
\begin{tabular}{l@{\extracolsep{-5mm}}l}
\textbf{\large Latin alphabet} \\[0.1cm]
$a$ & equation of state parameter \\
$a$ & interaction site index \\
$A$ & Helmholtz energy \\
$b$ & equation of state parameter \\
$b$ & interaction site index \\
$c$ & equation of state parameter \\
$c$ & coefficient of correlation function \\
$C$ & coefficient of correlation function \\
$h$ & enthalpy \\
$i$ & molecule index \\
$j$ & molecule index \\
$J$ & integral over intermolecular potential \\
$k_B$ & Boltzmann constant \\
$L$ & molecular elongation \\
$m$ & chain length \\
$N$ & number of particles \\
$p$ & pressure \\
$Q$ & quadrupolar moment \\
$r$ & site-site distance \\
$r_{\rm c}$ & center-center cut-off radius \\
$T$ & temperature \\
$u$ & pair potential \\
$x$ & mole fraction in the saturated liquid \\
$y$ & mole fraction in the saturated vapor \\
$Z$ & compressibility \\
\end{tabular}

\vspace{1.0cm} \noindent
\begin{tabular}{l@{\extracolsep{-10mm}}l}
\textbf{\large Vector properties} \\[0.1cm]
$\bm r$ & position vector \\
$\bm \omega$ & orientation vector \\
\end{tabular}

\noindent
\begin{tabular}{l@{\extracolsep{-5mm}}l}
\textbf{\large Greek alphabet} \\[0.1cm]
$\alpha$ & equation of state parameter \\
$\Delta h_{\rm v}$ & enthalpy of vaporization \\
$\Delta t$ & integration time step\\
$\epsilon$ & Lennard-Jones energy parameter \\
$\epsilon_{\rm s}$ & relative permittivity of dielectric continuum \\
$\eta$ & packing fraction \\
$\theta$ & angle of nutation \\
$\mu$ & dipolar moment \\
$\rho$ & density \\
$\sigma$ & Lennard-Jones size parameter \\
$\phi_{ij}$ & azimuthal angle between the axis of molecules $i$ and $j$\\

$\omega$ & acentric factor \\
\end{tabular}

\vspace{0.2cm}
\noindent
\begin{tabular}{l@{\extracolsep{10mm}}l}
\textbf{\large Subscript} \\[0.1cm]
c & property at critical point \\
D & dipole \\
$i$ & component index \\
$j$ & component index \\
$k$ & component index \\
Q & quadrupole \\
$\sigma$ & vapor-liquid coexistence \\
2CLJ & two-center Lennard-Jones \\
2CLJD & two-center Lennard-Jones plus point dipole \\
2CLJQ & two-center Lennard-Jones plus point quadrupole \\
2CLJDQ & two-center Lennard-Jones plus point dipole and point quadrupole \\
\end{tabular}

\vspace{0.2cm}
\noindent
\begin{tabular}{ll}
\textbf{\large Superscript} \\[0.1cm]
*     & reduced property\\
$'$   & saturated liquid \\
$''$  & saturated vapor \\
id  & ideal gas property \\
res & residual property \\
TS & tangent sphere \\
\end{tabular}
\clearpage

\addcontentsline{toc}{section}{\ \ \ \ References}

\clearpage

\begin{table}[ht]
\noindent 
\caption[]{Vapor-liquid equilibrium data of twelve 2CLJDQ model fluids. 
The number in parentheses indicates the statistical uncertainty in the last decimal digit.}
\label{t2cljdqvle}
\medskip
\begin{center}
{\footnotesize
\begin{tabular}{c|ccccc}\hline\hline
{\normalsize $\!T^*$} & {\normalsize $p_{\sigma}^*$} & {\normalsize $\rho'^*$} & {\normalsize $\rho''^*$} & {\normalsize
$h'^{\mbox{\scriptsize res*}}$} & {\normalsize $h''^{\mbox{\scriptsize res*}}$} \\ \hline

\multicolumn{6}{l}{{\normalsize $L^*=0$, $\mu^{*2}=6$, $Q^{*2}=2$}} \\ \hline
3.78268 & 0.0059  (6) & 0.8943  (2)& 0.0013  (3)& -42.01  (1)& \- 0.2\0 (2)\\
4.12656 & 0.0152  (9) & 0.8510  (2)& 0.0038  (3)& -40.38  (1)& -0.5\0 (2)\\
4.47044 & 0.032\0 (1) & 0.8240  (3)& 0.0079  (4)& -38.74  (1)& -1.3\0 (2)\\
4.81432 & 0.059\0 (2) & 0.7870  (3)& 0.0138  (6)& -37.10  (1)& -1.8\0 (1)\\
5.15820 & 0.104\0 (2) & 0.7463  (4)& 0.0237  (6)& -35.35  (2)& -2.60  (7)\\
5.50208 & 0.170\0 (2) & 0.6997  (4)& 0.0386  (5)& -33.42  (2)& -3.85  (7)\\
5.84596 & 0.260\0 (2) & 0.6474  (8)& 0.0605  (7)& -31.29  (3)& -5.41  (9)\\
6.18984 & 0.381\0 (3) & 0.584\0 (1)& 0.097\0 (2)& -28.74  (5)& -7.8\0 (2)\\
6.31849 & 0.437\0 (4) & 0.552\0 (2)& 0.116\0 (3)& -27.51  (7)& -8.9\0 (2)\\ \hline

\multicolumn{6}{l}{{\normalsize $L^*=0$, $\mu^{*2}=6$, $Q^{*2}=4$}} \\ \hline
4.18204 & 0.005  (2) & 0.96057  (3)& 0.0013  (4) &-54.11  (2)  &\0 -0.5\0  (1)\\
4.59192 & 0.008  (2) & 0.92333  (3)& 0.0018  (5) &-51.84  (2)  &\0 -0.5\0  (1)\\
4.97458 & 0.023  (2) & 0.88729  (2)& 0.0050  (3) &-49.70  (1)  &\0 -1.18   (3)\\
5.35724 & 0.044  (3) & 0.84937  (3)& 0.0090  (7) &-47.57  (1)  &\0 -1.8\0  (2)\\
5.73990 & 0.080  (3) & 0.80792  (4)& 0.0157  (7) &-45.32  (2)  &\0 -2.6\0  (1)\\
6.12256 & 0.137  (3) & 0.76301  (4)& 0.0277  (9) &-42.97  (2)  &\0 -4.2\0  (1)\\
6.50522 & 0.227  (3) & 0.71269  (6)& 0.0469  (1) &-40.42  (3)  &\0 -5.8\0  (2)\\
6.84334 & 0.320  (6) & 0.6582\0 (2)& 0.067\0 (2) &-37.75  (8)  &\0 -7.6\0  (2)\\
7.03343 & 0.394  (5) & 0.6257\0 (2)& 0.085\0 (3) &-36.25  (5)  &\0 -8.9\0  (3)\\
7.22352 & 0.477  (5) & 0.5852\0 (2)& 0.112\0 (3) &-34.33  (9)  &\- -11.0\0 (3)\\ \hline

\multicolumn{6}{l}{{\normalsize $L^*=0$, $\mu^{*2}=12$, $Q^{*2}=2$}} \\ \hline
4.36466 &  0.0049   (5) & 0.9208   (3) & 0.0012   (1)  &-53.68  (1) &\0 -0.62   (6)\\
4.76145 &  0.007\0  (1) & 0.8868   (3) & 0.0015   (2)  &-51.82  (1) &\0 -0.6\0   (1)\\
5.32623 &  0.029\0  (2) & 0.8367   (3) & 0.0060   (4)  &-49.18  (1) &\0 -1.8\0  (2)\\
5.73594 &  0.052\0  (2) & 0.7971   (3) & 0.0102   (4)  &-47.19  (2) &\0 -2.51   (8)\\
6.14565 &  0.096\0  (2) & 0.7556   (4) & 0.0188   (5)  &-45.15  (2) &\0 -3.9\0  (1)\\
6.55536 &  0.161\0  (3) & 0.7088   (5) & 0.0318   (6)  &-42.91  (2) &\0 -5.6\0  (1)\\
6.96507 &  0.249\0  (3) & 0.6550   (7) & 0.0509   (9)  &-40.41  (3) &\0 -7.7\0  (1)\\
7.37478 &  0.380\0  (4) & 0.590\0  (1) & 0.087\0  (2)  &-37.42  (4) &\- -11.2\0 (2)\\
7.53896 &  0.432\0  (4) & 0.554\0  (2) & 0.100\0  (2)  &-35.81  (8) &\- -12.1\0 (2)\\
7.57963 &  0.445\0  (5) & 0.544\0  (2) & 0.106\0  (3)  &-35.40  (7) &\- -12.7\0 (3)\\ \hline
\end{tabular}}
\end{center}
\end{table}
\clearpage

\begin{table}[ht]
\noindent
\begin{center}
Table \ref{t2cljdqvle}: continued. \\
\end{center}\medskip
\begin{center}
{\footnotesize
\begin{tabular}{c|ccccc}\hline
{\normalsize $\!T^*$} & {\normalsize $p_{\sigma}^*$} & {\normalsize $\rho'^*$} & {\normalsize $\rho''^*$} & {\normalsize
$h'^{\mbox{\scriptsize res*}}$} & {\normalsize $h''^{\mbox{\scriptsize res*}}$} \\ \hline \multicolumn{6}{l}{{\normalsize
$L^*=0$, $\mu^{*2}=12$, $Q^{*2}=4$}} \\ \hline
5.25366  & 0.003  (3) &0.9433   (3) &0.0006   (5)  &-63.53  (2)  &\0 -0.4  (3)\\
5.69146  & 0.015  (2) &0.9075   (3) &0.0027   (4)  &-61.18  (2)  &\0 -1.4  (2)\\
6.12927  & 0.030  (4) &0.8696   (3) &0.0053   (9)  &-58.80  (1)  &\0 -2.1  (5)\\
6.56707  & 0.057  (3) &0.8296   (3) &0.0099   (7)  &-56.38  (2)  &\0 -3.3  (2)\\
7.00488  & 0.103  (6) &0.7858   (5) &0.018\0  (1)  &-53.81  (2)  &\0 -4.5  (4)\\
7.44268  & 0.185  (4) &0.7379   (6) &0.034\0  (1)  &-51.08  (3)  &\0 -7.5  (3)\\
7.88049  & 0.290  (5) &0.6834   (9) &0.049\0  (2)  &-48.07  (4)  &\- -10.1  (2)\\
8.09939  & 0.347  (6) &0.651\0  (1) &0.060\0  (4)  &-46.33  (4)  &\- -10.9  (6)\\
8.45000  & 0.478  (8) &0.591\0  (2) &0.104\0  (4)  &-43.21  (7)  &\- -15.2  (5)\\ \hline

\multicolumn{6}{l}{{\normalsize $L^*=0.505$, $\mu^{*2}=6$, $Q^{*2}=2$}} \\ \hline
1.94604  & 0.0016   (3) &0.5765   (3)&0.0008   (2) &-23.96\0  (1)  &-0.21   (5)\\
2.12295  & 0.0028   (5) &0.5542   (3)&0.0014   (2) &-22.96\0  (1)  &-0.30   (4)\\
2.29986  & 0.0064   (7) &0.5313   (2)&0.0030   (3) &-21.979   (9)  &-0.48   (6)\\
2.47678  & 0.0142   (8) &0.5076   (2)&0.0062   (4) &-21.011   (9)  &-0.80   (5)\\

2.65369  & 0.026\0  (2) &0.4816   (6)&0.0113   (9) &-19.99\0  (2)  &-1.3\0  (1)\\
2.83060  & 0.047\0  (2) &0.4525   (4)&0.021\0  (1) &-18.88\0  (1)  &-2.2\0  (3)\\
3.00752  & 0.074\0  (1) &0.4184   (6)&0.032\0  (1) &-17.64\0  (2)  &-2.7\0  (2)\\
3.18443  & 0.110\0  (1) &0.377\0  (1)&0.050\0  (1) &-16.20\0  (4)  &-4.0\0  (2)\\
3.27289  & 0.131\0  (1) &0.351\0  (4)&0.064\0  (2) &-15.29\0  (9)  &-4.7\0  (1)\\ \hline

\multicolumn{6}{l}{{\normalsize $L^*=0.505$, $\mu^{*2}=6$, $Q^{*2}=4$}} \\ \hline
2.30000  & 0.0011   (5)&0.5893   (2)&0.0005   (2)   &-28.78  (1) &-0.2\0  (1)\\
2.49167  & 0.0056   (7)&0.5664   (2)&0.0024   (3)   &-27.53  (1) &-0.67   (7)\\
2.68333  & 0.009\0  (2)&0.5426   (3)&0.0037   (9)   &-26.30  (1) &-0.83   (2)\\
2.87500  & 0.017\0  (3)&0.5162   (3)&0.007\0  (1)   &-25.01  (1) &-1.2\0  (2)\\
3.06667  & 0.033\0  (4)&0.4884   (4)&0.013\0  (2)   &-23.69  (1) &-1.9\0  (3)\\
3.25833  & 0.058\0  (2)&0.4562   (4)&0.023\0  (1)   &-22.25  (2) &-2.8\0  (1)\\
3.45000  & 0.093\0  (2)&0.4198   (7)&0.039\0  (2)   &-20.66  (2) &-4.3\0  (2)\\
3.54583  & 0.117\0  (3)&0.3983   (9)&0.052\0  (2)   &-19.79  (3) &-5.1\0  (2)\\
3.64167  & 0.143\0  (2)&0.372\0  (1)&0.070\0  (2)   &-18.77  (4) &-6.4\0  (1)\\ \hline
\end{tabular}}
\end{center}
\end{table}
\clearpage

\begin{table}[ht]
\noindent
\begin{center}
Table \ref{t2cljdqvle}: continued. \\
\end{center}\medskip
\begin{center}
{\footnotesize
\begin{tabular}{c|ccccc}\hline
{\normalsize $\!T^*$} & {\normalsize $p_{\sigma}^*$} & {\normalsize $\rho'^*$} & {\normalsize $\rho''^*$} & {\normalsize
$h'^{\mbox{\scriptsize res*}}$} & {\normalsize $h''^{\mbox{\scriptsize res*}}$} \\ \hline

\multicolumn{6}{l}{{\normalsize $L^*=0.505$, $\mu^{*2}=12$, $Q^{*2}=2$}} \\ \hline
2.23188  & 0.0011   (3)&0.5917   (2)&0.0005   (2) &-30.816   (9)  &-0.63   (9)\\
2.44569  & 0.0016   (6)&0.5690   (2)&0.0007   (3) &-29.62\0   (1)  &-0.3\0  (3)\\
2.64949  & 0.0040   (6)&0.5466   (2)&0.0016   (3) &-28.47\0  (1)  &-0.74   (9)\\
2.85330  & 0.0093   (9)&0.5226   (3)&0.0036   (4) &-27.33\0  (1)  &-1.2\0  (2)\\
3.06000  & 0.0189   (7)&0.4964   (3)&0.0072   (3) &-26.10\0  (1)  &-1.88   (9)\\
3.26092  & 0.0304   (2)&0.4687   (4)&0.012\0  (1) &-24.87\0  (2)  &-2.9\0  (3)\\
3.46472  & 0.055\0  (1)&0.4384   (5)&0.0213   (6) &-23.57\0  (2)  &-3.76   (9)\\
3.66853  & 0.080\0  (2)&0.399\0  (1)&0.034\0  (1) &-21.97\0  (3)  &-5.0\0  (1)\\
3.85506  & 0.116\0  (3)&0.354\0  (2)&0.050\0  (3) &-20.20\0  (5)  &-6.4\0  (3)\\ \hline

\multicolumn{6}{l}{{\normalsize $L^*=0.505$, $\mu^{*2}=12$, $Q^{*2}=4$}} \\ \hline
2.47462  & 0.00052  (1)&0.6181   (2)&0.0001\0 (2) &-37.11  (1)  &-0.5\0  (4)\\
2.69958  & 0.00124  (1)&0.5958   (2)&0.00034  (5) &-35.67  (1)  &-0.1\0  (2)\\
2.92455  & 0.00322  (1)&0.5722   (3)&0.00110  (3) &-34.23  (1)  &-1.17   (6)\\
3.14951  & 0.00787  (2)&0.5477   (3)&0.00275  (1) &-32.79  (1)  &-1.54   (2)\\
3.37448  & 0.01649  (8)&0.5214   (3)&0.00562  (5) &-31.32  (1)  &-2.22   (3)\\
3.59944  & 0.0304\0  (1)&0.4926   (4)&0.01021  (7) &-29.82  (2)  &-3.11   (4)\\
3.82441  & 0.049\0\0  (3)&0.4593   (6)&0.0177\0  (6) &-28.10  (3)  &-4.5\0  (1)\\
4.04937  & 0.080\0\0  (4)&0.422\0  (1)&0.029\0\0 (3) &-26.29  (4)  &-5.8\0  (5)\\
4.16186  & 0.105\0\0  (4)&0.400\0  (1)&0.042\0\0 (3) &-25.28  (4)  &-7.2\0  (5)\\
4.27434  & 0.132\0\0  (3)&0.371\0  (2)&0.056\0\0 (2) &-23.99  (6)  &-8.4\0  (3)\\ \hline

\multicolumn{6}{l}{{\normalsize $L^*=1$, $\mu^{*2}=6$, $Q^{*2}=2$}} \\ \hline
1.35000 &0.00013  (1)  &0.4694  (2)  &0.00009   (3) &-23.03  (2) &-0.25  (1)\\
1.48000 &0.00060  (1)  &0.4466  (3)  &0.00035   (4) &-21.50  (2) &-0.29  (1)\\
1.60000 &0.00187  (3)  &0.4257  (3)  &0.00126   (5) &-20.22  (2) &-0.56  (6)\\
1.72000 &0.00448  (2)  &0.4036  (3)  &0.00280   (6) &-18.98  (1) &-0.93  (4)\\
1.85000 &0.00989  (3)  &0.3781  (2)  &0.00629   (3) &-17.63  (1) &-1.40  (2)\\
1.96000 &0.01775  (7)  &0.3549  (3)  &0.01138   (6) &-16.50  (1) &-1.93  (1)\\
2.09000 &0.0326\0  (2)  &0.3233   (5)   &0.0218\0  (2) &-15.06  (2) &-2.80  (4)\\
2.21000 &0.0522\0  (3)  &0.2840   (9)   &0.0370\0  (5) &-13.46  (3) &-3.86  (8)\\
2.26000 &0.0627\0  (4)  &0.251\0  (2)   &0.0456\0  (7) &-12.35  (5) &-4.34  (9)\\ \hline

\end{tabular}}
\end{center}
\end{table}
\clearpage

\begin{table}[ht]
\noindent
\begin{center}
Table \ref{t2cljdqvle}: continued. \\

\end{center}\medskip
\begin{center}
{\footnotesize
\begin{tabular}{c|ccccc}\hline
{\normalsize $\!T^*$} & {\normalsize $p_{\sigma}^*$} & {\normalsize $\rho'^*$} & {\normalsize $\rho''^*$} & {\normalsize
$h'^{\mbox{\scriptsize res*}}$} & {\normalsize $h''^{\mbox{\scriptsize res*}}$} \\ \hline

\multicolumn{6}{l}{{\normalsize $L^*=1$, $\mu^{*2}=6$, $Q^{*2}=4$}} \\ \hline

1.84000   &0.0005   (1)  &0.4787  (8)   &0.00033  (2) &-30.34  (8) &-1.0\0   (2)  \\
2.00000   &0.0019   (1)  &0.4505  (3)   &0.00100  (2) &-28.14  (2) &-1.38    (9)  \\
2.13000   &0.0047   (4)  &0.4280  (4)   &0.00263  (4) &-26.43  (3) &-2.12    (9)  \\
2.25000   &0.0101   (6)  &0.4052  (4)   &0.0055\0   (4) &-24.85  (3) &-2.9\0   (2)  \\
2.40000   &0.0238   (3)  &0.3731  (4)   &0.0136\0   (4) &-22.79  (2) &-4.4\0   (2)  \\
2.55000   &0.040\0  (1)  &0.3367  (6)   &0.0236\0   (1) &-20.68  (3) &-5.1\0   (2)  \\
2.60000   &0.049\0  (1)  &0.3251  (8)   &0.0305\0   (3) &-19.94  (3) &-5.8\0   (3)  \\
2.68000   &0.065\0  (5)  &0.300\0 (1)   &0.047\0\0  (8) &-18.70  (4) &-7.1\0   (8)  \\
2.75000   &0.091\0  (8)  &0.267\0 (2)   &0.07\0\0\0 (1) &-17.13  (7) &-9\0\0\0 (1)  \\ \hline

\multicolumn{6}{l}{{\normalsize $L^*=1$, $\mu^{*2}=12$, $Q^{*2}=2$}} \\ \hline
1.64000  &0.0001  (1)  &0.4773  (5)  &0.00006 (1) &-30.49  (4) &-1.1\0  (3)\\
1.80000  &0.0004  (1)  &0.4572  (3)  &0.00018 (2) &-29.02  (2) &-1.35   (9)\\
1.94000  &0.0011  (1)  &0.4350  (3)  &0.00065 (1) &-27.31  (3) &-1.76   (5)\\
2.10000  &0.0037  (3)  &0.4096  (3)  &0.0014\0 (3) &-25.52  (2) &-2.0\0  (4)\\
2.25000  &0.0088  (5)  &0.3820  (3)  &0.0046\0 (6) &-23.75  (2) &-3.2\0  (3)\\
2.37000  &0.0148  (1)  &0.3588  (4)  &0.00853 (9) &-22.36  (2) &-4.03   (4)\\
2.50000  &0.0276  (5)  &0.3305  (5)  &0.0173\0  (5) &-20.77  (2) &-5.3\0  (1)\\
2.70000  &0.0543  (5)  &0.264\0 (2)  &0.0353\0  (7) &-17.64  (6) &-6.9\0  (1)\\
2.80000  &0.073\0 (1)  &0.228\0 (3)  &0.054\0\0 (1) &-16.01  (9) &-7.9\0  (1)\\ \hline

\multicolumn{6}{l}{{\normalsize $L^*=1$, $\mu^{*2}=12$, $Q^{*2}=4$}} \\ \hline
2.14000   &0.00010 (1)  &0.4897  (9)   &0.00008 (1) &-40.21  (7) &\0 -3.2\0 (3)\\
2.31000   &0.00051 (1)  &0.4790  (7)   &0.00032 (1) &-39.43  (7) &\0 -4.2\0 (2)\\
2.48000   &0.00157 (2)  &0.461\0 (1)   &0.00082 (3) &-38.0\0 (1) &\0 -4.4\0 (3)\\
2.61000   &0.00393 (7)  &0.4338  (6)   &0.00191 (3) &-35.64  (4) &\0 -5.56  (5)\\
2.75000   &0.0081\0  (2)  &0.4090  (6)   &0.0041\0   (1) &-33.55  (4) &\0 -6.8\0 (1)\\
2.97000   &0.023\0\0 (1)  &0.3646  (6)   &0.0125\0   (9) &-30.24  (3) &\0 -8.8\0 (3)\\
3.08000   &0.032\0\0 (1)  &0.3411  (8)   &0.0177\0   (9) &-28.60  (4) &\0 -9.4\0 (2)\\
3.19000   &0.047\0\0 (2)  &0.3132  (9)   &0.027\0\0  (2) &-26.79  (4) &\- -10.3\0 (4)\\ \hline\hline
\end{tabular}}
\end{center}
\end{table}
\clearpage

\begin{table}[ht]
\noindent 
\caption[]{Critical temperature, density, pressure and compressibility as well as acentric factor of twelve 2CLJDQ model fluids.}
\label{2CLJDQcdomtab}
\bigskip
\begin{center}
\begin{tabular}{|c|c||l|l|l|l|l|l|c|}\hline
\multicolumn{2}{|c||}{ } & \multicolumn{6}{|l|}{$L^*$} & \multicolumn{1}{||c|}{ } \\ \cline{3-8} \multicolumn{2}{|c||}{ } &
\multicolumn{2}{|c|}{$0$} &  \multicolumn{2}{|c|}{$0.505$} & \multicolumn{2}{|c|}{$1$} & \multicolumn{1}{||c|}{}\\
\cline{3-8} \multicolumn{2}{|c||}{ } & \multicolumn{6}{|l|}{$\mu^{*2}$}& \multicolumn{1}{||c|}{ }  \\ \cline{3-8}
\multicolumn{2}{|c||}{ } & \multicolumn{1}{|c|}{$6$}& \multicolumn{1}{|c|}{$12$}& \multicolumn{1}{|c|}{$6$}& \multicolumn{1}{|c|}{$12$} & \multicolumn{1}{|c|}{$6$}& \multicolumn{1}{|c|}{$12$}& \multicolumn{1}{||c|}{ } \\
\hline \hline

$Q^{*2}$ &   & \phantom{-}6.651    & \phantom{-}7.937     & \phantom{-}3.445     & \phantom{-}4.067     & \phantom{-}2.344      & \phantom{-}2.863    & \multicolumn{1}{||c|}{$T^*_{\rm c}$}\\ \cline{3-9}

         &   & \phantom{-}0.319    & \phantom{-}0.3128    & \phantom{-}0.1982    & \phantom{-}0.1929    & \phantom{-}0.1508     & \phantom{-}0.1350   & \multicolumn{1}{||c|}{$\rho^*_{\rm c}$} \\ \cline{3-9}
         & 2 & \phantom{-}0.5995   & \phantom{-}0.6125    & \phantom{-}0.1844    & \phantom{-}0.1692    & \phantom{-}0.0822     & \phantom{-}0.0897   & \multicolumn{1}{||c|}{$p^*_{\rm c}$} \\ \cline{3-9}
         &   & \phantom{-}0.2823   & \phantom{-}0.2467    & \phantom{-}0.2701    & \phantom{-}0.2157    & \phantom{-}0.2324     & \phantom{-}0.2320   & \multicolumn{1}{||c|}{$Z_{\rm c}$} \\ \cline{3-9}
         &   & \phantom{-}0.1254   & \phantom{-}0.1733    & \phantom{-}0.1987    & \phantom{-}0.2423    & \phantom{-}2.4890     & \phantom{-}2.8628   & \multicolumn{1}{||c|}{$\omega$} \\ \cline{2-9}
         &   & \phantom{-}7.604    & \phantom{-}8.944     & \phantom{-}3.832     & \phantom{-}4.501     & \phantom{-}2.861      & \phantom{-}3.424    & \multicolumn{1}{||c|}{$T^*_{\rm c}$} \\ \cline{3-9}
         &   & \phantom{-}0.3288   & \phantom{-}0.3229    & \phantom{-}0.2086    & \phantom{-}0.1989    & \phantom{-}0.1532     & \phantom{-}0.1481   & \multicolumn{1}{||c|}{$\rho^*_{\rm c}$} \\ \cline{3-9}
         & 4 & \phantom{-}0.6797   & \phantom{-}0.7036    & \phantom{-}0.2121    & \phantom{-}0.2382    & \phantom{-}0.1335     & \phantom{-}0.1112   & \multicolumn{1}{||c|}{$p^*_{\rm c}$} \\ \cline{3-9}
         &   & \phantom{-}0.2718   & \phantom{-}0.2436    & \phantom{-}0.2654    & \phantom{-}0.2660    & \phantom{-}0.3045     & \phantom{-}0.2191   & \multicolumn{1}{||c|}{$Z_{\rm c}$} \\ \cline{3-9}
         &   & \phantom{-}0.2057   & \phantom{-}0.2783    & \phantom{-}0.3296    & \phantom{-}2.4341    & \phantom{-}3.2371     & \phantom{-}3.8008   & \multicolumn{1}{||c|}{$\omega$} \\ \hline
\end{tabular}
\end{center}
\end{table}
\clearpage

\begin{table}[ht]
\noindent
\caption[]{Model constants of the new dipole-quadrupole EOS contribution. In Eq. (\ref{A3dq}), the parameter $\alpha$=1.19374.}
\label{modelconstants}
\medskip
\begin{center}
{\footnotesize
\begin{tabular}{c|cccccccc}\hline\hline
{\normalsize $i$} & {\normalsize $a_{0i}$} & {\normalsize $a_{1i}$} & {\normalsize $a_{2i}$} & {\normalsize
$b_{0i}$} & {\normalsize $b_{1i}$} & {\normalsize $b_{2i}$} & {\normalsize $c_{0i}$} & {\normalsize $c_{1i}$}\\ \hline
0 & 0.6970950 & -0.6734593 & 0.6703408 & -0.4840383 & 0.6765101 & -1.1675601 & 7.846431 &-20.72202\\
1 & -0.6335541 & -1.4258991 & -4.3384718 & 1.9704055 & -3.0138675 & 2.1348843 & 33.42700 & -58.63904\\
2 & 2.9455090 & 4.1944139 & 7.2341684 & -2.1185727 & 0.4674266 & 0 & 4.689111 & -1.764887\\
3 & -1.4670273 & 1.0266216 & 0 & 0 & 0 & 0 & 0 & 0\\ \hline\hline
\end{tabular}}
\end{center}
\end{table}
\clearpage

\begin{table}[ht]
\noindent
\caption[]{Binary interaction parameters taken from \cite{huang}.}
\label{xi}
\medskip
\begin{center}
\begin{tabular}{l|c}\hline\hline
Mixture & $\xi$\\ \hline
C$_2$H$_2$+R152a & 1.090 \\
R142b+R113       & 0.952 \\
CO$_2$+R12       & 0.927 \\
R22+R142b        & 0.985 \\
Propylene+R115   & 0.948 \\ \hline\hline
\end{tabular}
\end{center}
\end{table}
\clearpage

\begin{table}[ht]
\noindent 
\caption[]{Vapor-liquid equilibrium data for five binary mixtures taken from \cite{huang}.
Pure substance vapor pressures were obtained via correlations provided in \cite{stoll4549,stoll179}.
The number in parentheses indicates the statistical uncertainty in the last decimal digit.}

\label{vlemix}
\medskip
\begin{center}
\begin{tabular}{l|ll}\hline\hline
$x_1$ / mol/mol & $p$ / MPa & $y_1$ / mol/mol \\ \hline
\multicolumn{3}{l}{C$_2$H$_2$+R152a (1+2) at 303.2 K}          \\ \hline
0     & 0.69     & 0          \\[-0.11cm]
0.128 & 0.87 (5) & 0.32\0 (2) \\[-0.11cm]
0.424 & 1.77 (5) & 0.76\0 (2) \\[-0.11cm]
0.569 & 2.45 (8) & 0.88\0 (2) \\[-0.11cm]
0.801 & 4.01 (6) & 0.950 (4)  \\[-0.11cm]
1     & 5.62     & 1          \\ \hline
\multicolumn{3}{l}{R142b+R113 (1+2) at 373 K}                \\ \hline
0     & 0.45     & 0          \\[-0.11cm]
0.252 & 0.90 (3) & 0.57\0 (2) \\[-0.11cm]
0.502 & 1.27 (4) & 0.77\0 (2) \\[-0.11cm]
0.751 & 1.63 (4) & 0.890 (5)  \\[-0.11cm]
1     & 2.00     & 1          \\ \hline
\multicolumn{3}{l}{CO$_2$+R12 (1+2) at 273 K}                \\ \hline
0     & 0.30     & 0          \\[-0.11cm]
0.147 & 0.86 (2) & 0.65\0 (1) \\[-0.11cm]
0.388 & 1.70 (2) & 0.849 (6)  \\[-0.11cm]
0.550 & 2.20 (2) & 0.899 (4)  \\[-0.11cm]
0.714 & 2.67 (2) & 0.932 (4)  \\[-0.11cm]
1     & 3.52     & 1          \\ \hline
\multicolumn{3}{l}{R22+R142b (1+2) at 328.15 K}                 \\ \hline
0     & 0.75     & 0          \\[-0.11cm]
0.218 & 1.02 (1) & 0.40\0 (1) \\[-0.11cm]
0.455 & 1.36 (1) & 0.66\0 (1) \\[-0.11cm]
0.560 & 1.50 (3) & 0.730 (8)  \\[-0.11cm]
0.661 & 1.66 (2) & 0.806 (5)  \\[-0.11cm]
0.806 & 1.86 (2) & 0.900 (3)  \\[-0.11cm]
1     & 2.17     & 1          \\ \hline
\multicolumn{3}{l}{Propylene+R115 (1+2) at 298 K}            \\ \hline
0     & 0.95     & 0          \\[-0.11cm]
0.113 & 0.98 (4) & 0.175 (7)  \\[-0.11cm]
0.316 & 1.18 (3) & 0.386 (8)  \\[-0.11cm]
0.549 & 1.24 (2) & 0.59\0 (1) \\[-0.11cm]
0.788 & 1.26 (2) & 0.780 (7)  \\[-0.11cm]
1     & 1.19     & 1          \\ \hline\hline
\end{tabular}
\end{center}
\end{table}
\clearpage

\listoffigures

\clearpage

\clearpage

\clearpage

\clearpage

\clearpage

\clearpage

\clearpage

\clearpage

\clearpage

\clearpage

\clearpage

\begin{figure}[ht]
\begin{center}
\epsfig{file=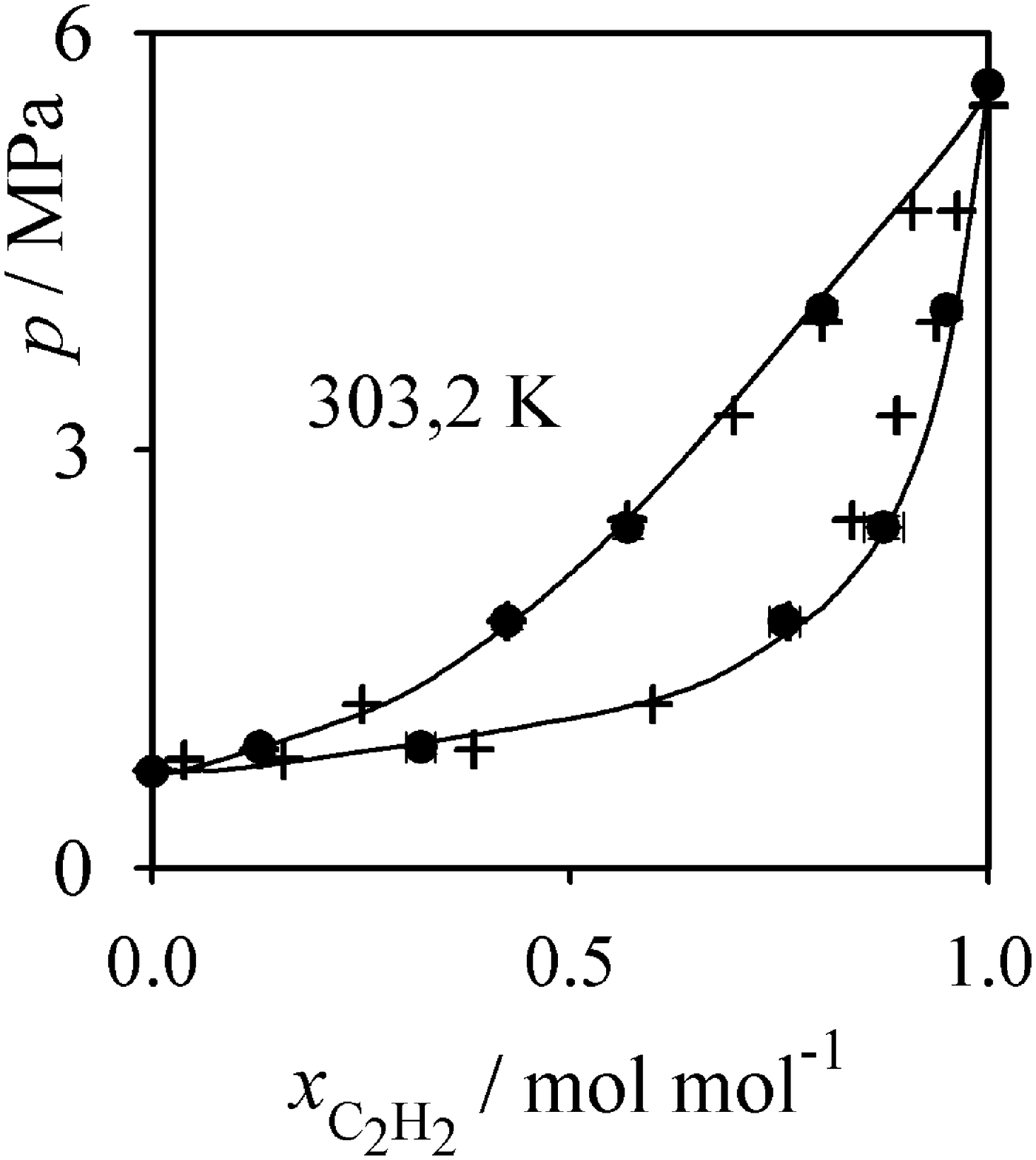, scale=0.47}
\end{center}
\vskip1cm
\caption[Vapor-liquid phase diagram of C$_2$H$_2$+R152a at 303.2 K: 
{\bf -----}~present EOS, 
{\large $\bullet$}~simulation data \cite{huang}, $+$~experimental data  
\cite{expC2H2+R152A}.]{Vrabec and Gross}
\label{C2H2R152A}
\end{figure}
\clearpage

\begin{figure}[ht]
\begin{center}
\epsfig{file=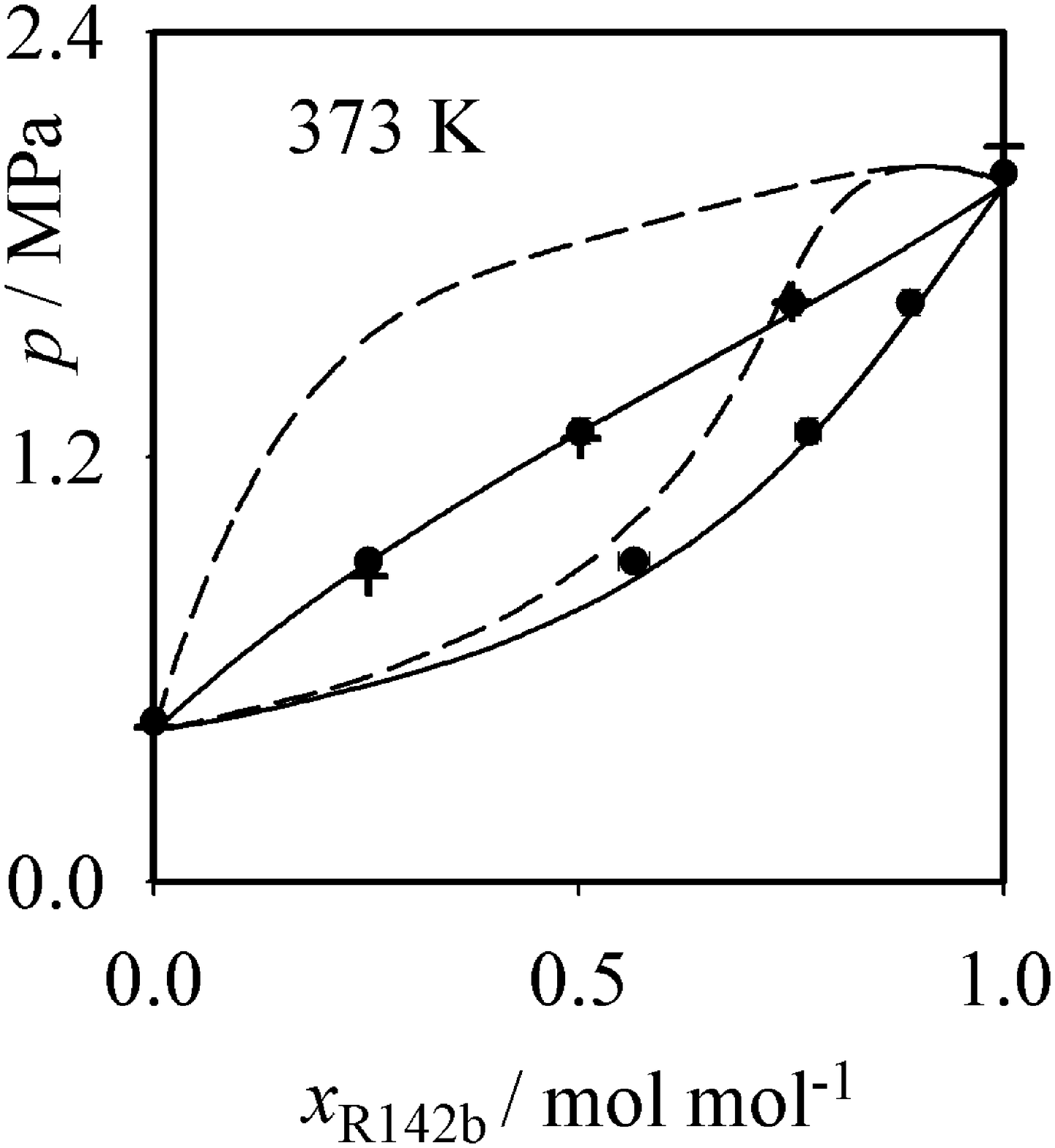, scale=0.47}
\end{center}
\vskip1cm
\caption[Vapor-liquid phase diagram of R142b+R113 at 373 K: 
{\bf -----}~present EOS, {\bf - - -}~present EOS neglecting
dipole-quadrupole cross-interactions, {\bf $\cdots$}~EOS proposed
by Weingerl and Fischer \cite{weingerl202},
{\large $\bullet$}~simulation data \cite{huang}, $+$~experimental data  
\cite{expR142B+R113}.]{Vrabec and Gross}
\label{R142BR113}
\end{figure}
\clearpage

\begin{figure}[ht]
\begin{center}
\epsfig{file=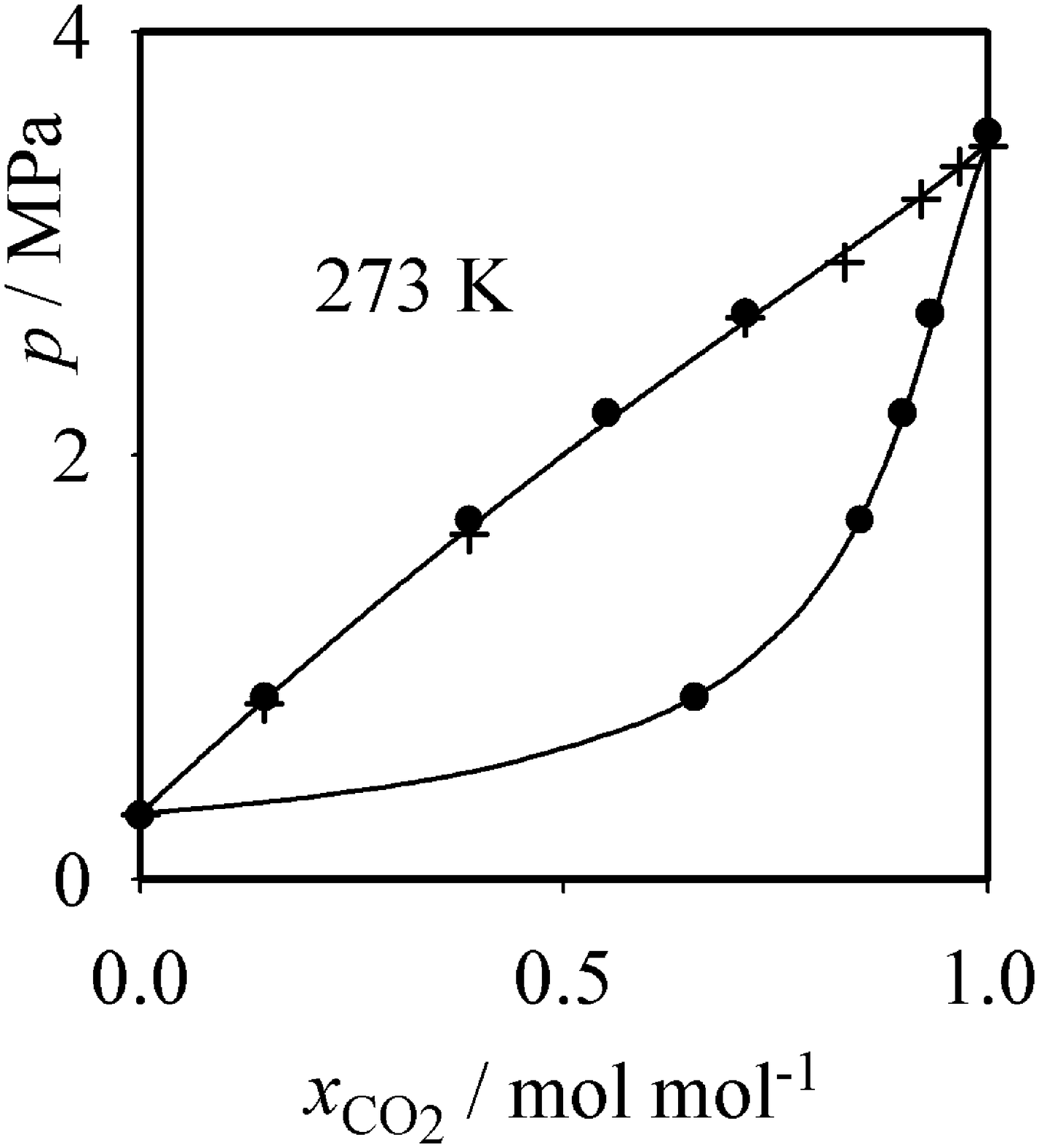, scale=0.47}
\end{center}
\vskip1cm
\caption[Vapor-liquid phase diagram of CO$_2$+R12 at 273 K: 
{\bf -----}~present EOS, 
{\large $\bullet$}~simulation data \cite{huang}, $+$~experimental data  
\cite{expCO2+R12}.]{Vrabec and Gross}
\label{CO2R12}
\end{figure}
\clearpage

\begin{figure}[ht]
\begin{center}
\epsfig{file=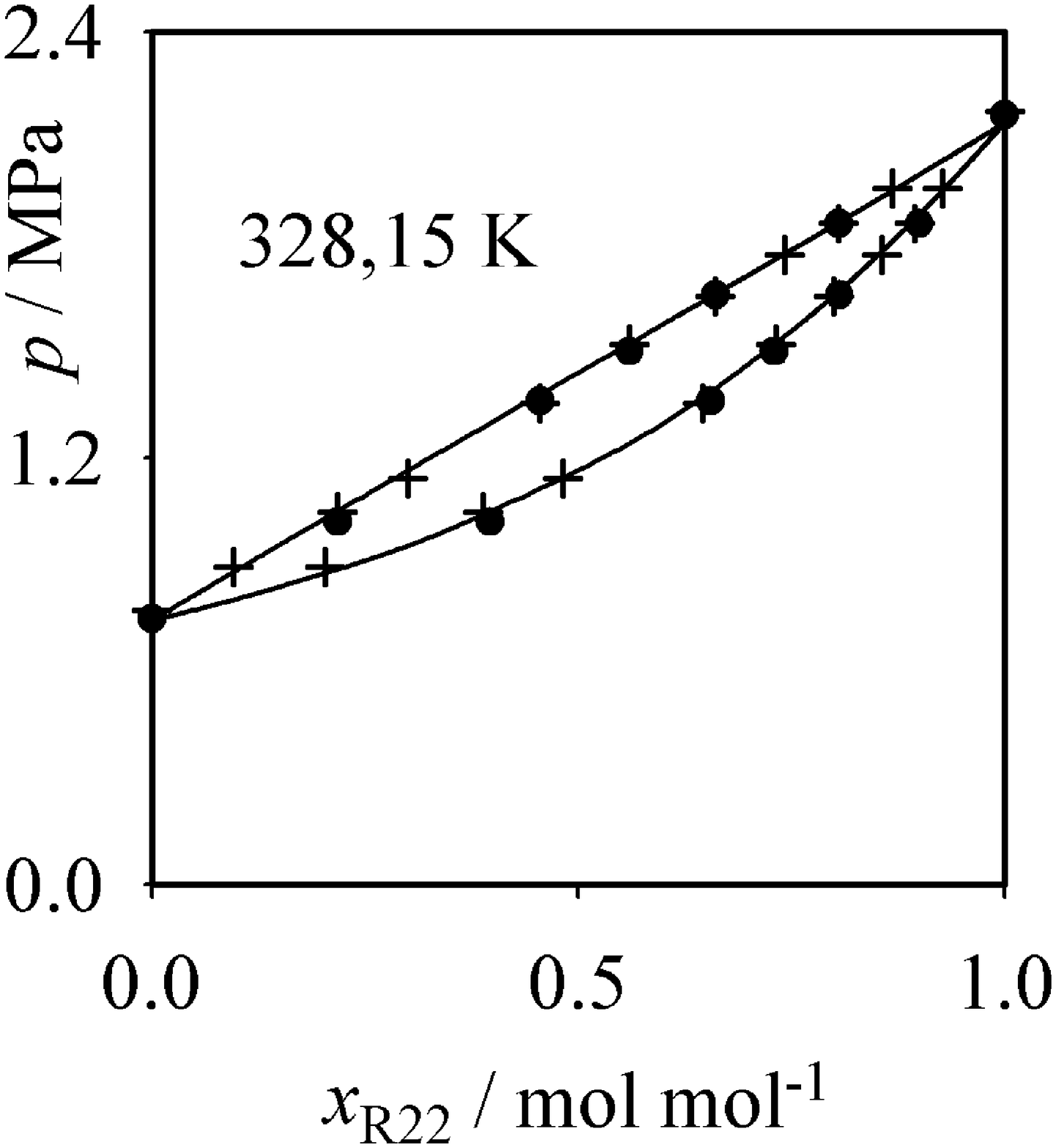, scale=0.47}
\end{center}
\vskip1cm
\caption[Vapor-liquid phase diagram of R22+R142b at 328.15 K: 
{\bf -----}~present EOS, 
{\large $\bullet$}~simulation data \cite{huang}, $+$~experimental data  
\cite{expR22+R142B}.]{Vrabec and Gross}
\label{R22R142B}
\end{figure}
\clearpage

\begin{figure}[ht]
\begin{center}
\epsfig{file=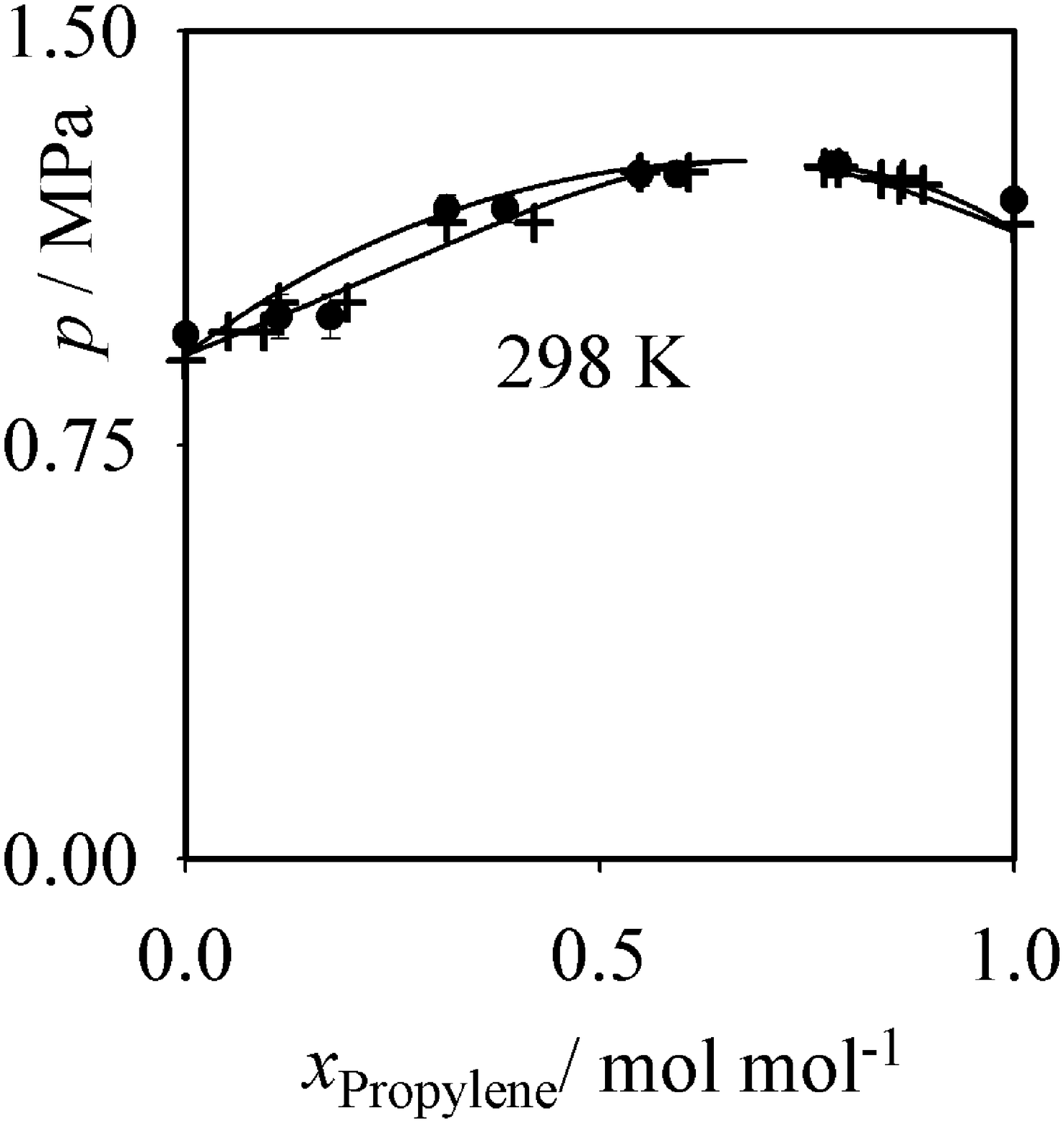, scale=0.47}
\end{center}
\vskip1cm
\caption[Vapor-liquid phase diagram of Propylene+R115 at 298 K: 
{\bf -----}~present EOS, 
{\large $\bullet$}~simulation data \cite{huang}, $+$~experimental data  
\cite{expPropylene+R115}.]{Vrabec and Gross}
\label{PropyleneR115}
\end{figure}
\clearpage


\begin{thebibliography}{99}

\bibitem{stoll4549} Stoll, J.; Vrabec, J.; Hasse, H.; Fischer, J. {\it Fluid Phase Equilib.} {\bf 2001}, {\it 179}, 339.
\bibitem{stoll179} Stoll, J.; Vrabec, J.; Hasse, H. {\it Fluid Phase Equilib.} {\bf 2003}, {\it 209}, 29.
\bibitem{stoll03} Stoll, J.; Vrabec, J.; Hasse, H. {\it J. Chem. Phys.} {\bf 2003}, {\it 119}, 11396.
\bibitem{vrabec105} Vrabec, J.; Stoll, J.; Hasse, H. {\it J. Phys. Chem. B} {\bf 2001}, {\it 105}, 12126.
\bibitem{stoll000} Stoll, J.; Vrabec, J.; Hasse, H. {\it AIChE J.} {\bf 2003}, {\it 49}, 2187.
\bibitem{vrabec05} Vrabec, J.; Stoll, J.; Hasse, H. {\it Mol. Sim.} {\bf 2005}, {\it 31}, 215.
\bibitem{stolldiss} Stoll, J. {\it Molecular Models for the Prediction of Thermophysical Properties of Pure Fluids and Mixtures}; Fortschritt-Berichte VDI, Reihe 3, 836, VDI Verlag: D\"usseldorf, 2005.
\bibitem{gross06} Gross, J.; Vrabec, J. {\it AIChE J.} {\bf 2006}, {\it 52}, 1194.
\bibitem{gross05} Gross, J. {\it AIChE J.} {\bf 2005}, {\it 51}, 2556.
\bibitem{dubey9421} Dubey, G. S.; O'Shea, S. F. {\it Phys. Rev. E} {\bf 1994}, {\it 49}, 2175.
\bibitem{stell1} Stell, G.; Rasaiah, J. C.; Narang, H. {\it Mol. Phys.} {\bf 1972}, {\it 23}, 393.
\bibitem{stell2} Stell, G.; Rasaiah, J. C.; Narang, H. {\it Mol. Phys.} {\bf 1974}, {\it 127}, 393.
\bibitem{rush1} Rushbrooke, G. S.; Stell, G.; Hoye, J. S. {\it Mol. Phys.} {\bf 1973}, {\it 26}, 1199.
\bibitem{hen1}   Henderson, D.; Blum, L.; Tani, A. {\it Equation of state of ionic fluids}; in: Chao, K. C.; Robinson, R. L.; Eds. {\it Equations of State. Theories and Applications}; ACS Symposium Series 300: Washington, DC, American Chemical Society, 1986, p. 281.
\bibitem{gub1} Gubbins, K. E.; Twu, C. H. {\it Chem. Eng. Sci.} {\bf 1978}, {\it 33}, 863.
\bibitem{luc1} Luckas, M.; Lucas, K.; Deiters, U.; Gubbins, K. E. {\it Mol. Phys.} {\bf 1986}, {\it 57}, 241.
\bibitem{mos1} Moser, B.; Lucas, K.; Gubbins, K. E. {\it Fluid Phase Equilib.} {\bf 1981}, {\it 7}, 153.
\bibitem{shu1} Shukla, K. P.; Lucas, K.; Moser, B. {\it Fluid Phase Equilib.} {\bf 1983}, {\it 15}, 125.
\bibitem{shu2} Shukla, K. P.; Lucas, K.; Moser, B. {\it Fluid Phase Equilib.} {\bf 1984}, {\it 17}, 19.
\bibitem{shu3} Shukla, K. P.; Lucas, K.; Moser, B. {\it Fluid Phase Equilib.} {\bf 1983}, {\it 15}, 125.
\bibitem{bou1} Boublik, T. {\it Mol. Phys.} {\bf 1992}, {\it 76}, 327.
\bibitem{saager9127} Saager, B.; Fischer, J.; Neumann, M. {\it Mol. Sim.} {\bf 1991}, {\it 6}, 27.
\bibitem{saager1992} Saager, B.; Fischer, J.; {\it Fluid Phase Equilib.} {\bf 1992}, {\it 72}, 67
\bibitem{weingerl202} Weingerl, U.; Fischer, J. {\it Fluid Phase Equilib.} {\bf 2002}, {\it 202}, 49.
\bibitem{chapman88} Chapman, W. G.; Jackson, G.; Gubbins, K. E. {\it Mol. Phys.} {\bf 1988}, {\it 65}, 1057.
\bibitem{wertheim8419} Wertheim, M. S. {\it J. Stat. Phys.} {\bf 1984}, {\it 35}, 19.
\bibitem{wertheim8435} Wertheim, M. S. {\it J. Stat. Phys.} {\bf 1984}, {\it 35}, 35.
\bibitem{wertheim86459} Wertheim, M. S. {\it J. Stat. Phys.} {\bf 1986}, {\it 42}, 459.
\bibitem{wertheim86477} Wertheim, M. S. {\it J. Stat. Phys.} {\bf 1986}, {\it 42}, 477.
\bibitem{bla1} Blas, F. J.; Vega, L. F. {\it Ind. Eng. Chem. Res.} {\bf 1998}, {\it 37}, 660.
\bibitem{bla2} Blas, F. J.; Vega, L. F. {\it J. Chem. Phys.} {\bf 1998}, {\it 109}, 7405.
\bibitem{pam1} P\'amies, C. J.; Vega, L. F. {\it Ind. Eng. Chem. Res.} {\bf 2001}, {\it 40}, 2532.
\bibitem{gil1} Gil-Villegas, A.; Galindo, A.; Whitehead, P. J.; Mills, S. J.; Jackson, G.; Burgess, A. N. {\it J. Chem. Phys.} {\bf 1997}, {\it 106}, 4168.
\bibitem{dav1} Davies, L. A.; Gil-Villegas, A.; Jackson, G. {\it Int. J. Thermophys.} {\bf 1998}, {\it 19}, 675.
\bibitem{gross00} Gross, J.; Sadowski, G. {\it Fluid Phase Equilib.} {\bf 2000}, {\it 168}, 183.
\bibitem{gross01} Gross, J.; Sadowski, G. {\it Ind. Eng. Chem. Res.} {\bf 2001}, {\it 40}, 1244.
\bibitem{kra1} Kraska, T.; Gubbins, K. E. {\it Ind. Eng. Chem. Res.} {\bf 1996}, {\it 35}, 4727.
\bibitem{jog1} Jog, P. K.; Chapman, W. G. {\it Mol. Phys.} {\bf 1999}, {\it 97}, 307.
\bibitem{jog2} Jog, P. K.; Sauer, S. G.; Blaesing, J.; Chapman, W. G. {\it Ind. Eng. Chem. Res.} {\bf 2001}, {\it 40}, 4641.
\bibitem{tum1} Tumakaka, F.; Sadowski, G. {\it Fluid Phase Equilib.} {\bf 2004}, {\it 217}, 233.
\bibitem{li1}  Li, X. S.; Lu, J. F.; Li, Y. G.; Liu, J. C. {\it Fluid Phase Equilib.} {\bf 1998}, {\it 153}, 215.
\bibitem{liu1} Liu, W. B.; Li, Y. G.; Lu, J. F. {\it Fluid Phase Equilib.} {\bf 1999}, {\it 158}, 595.
\bibitem{kar1} Karakatsani, E. K.; Spyriouni, T.; Economou, I. G. {\it AIChE J.} {\bf 2005}, {\it 51}, 2328.
\bibitem{kar2} Karakatsani, E. K.; Economou, I. G. {\it J. Phys. Chem. B} {\bf 2006}, {\it 110}, 9252.
\bibitem{zha1} Zhao, H. G.; McCabe, C. {\it J. Chem. Phys.} {\bf 2006}, {\it 125}, 104504.
\bibitem{kle1} Kleiner, M.; Gross, J. {\it AIChE J.} {\bf 2005}, {\it 51}, 2556.
\bibitem{wer1} Wertheim, M. S. {\it Mol. Phys.} {\bf 1977}, {\it 35}, 1109.
\bibitem{wer2} Wertheim, M. S. {\it Mol. Phys.} {\bf 1979}, {\it 37}, 83.
\bibitem{wer3} Wertheim, M. S. {\it Annu. Rev. Phys. Chem.} {\bf 1979}, {\it 30},  471.
\bibitem{gra1} Gray, C. G.; Joslin, C. G.; Venkatasubramanian, V.; Gubbins, K. E. {\it Mol. Phys.} {\bf 1985}, {\it 54}, 1129.
\bibitem{leon07} Leonhard, K.; Nguyen, V. N.; Lucas, K. {\it Fluid Phase Equilib. (Submitted, 2007)}
\bibitem{allen87} Allen, M. P.; Tildesley, D. J. {\it Computer Simulation of Liquids}; Clarendon Press: Oxford, 1987.
\bibitem{widom6328}  Widom, B. {\it J. Chem. Phys.} {\bf 1963}, {\it 39}, 2808.
\bibitem{moeller9435} M\"oller, D.; Fischer, J. {\it Fluid Phase Equilib.} {\bf 1994}, {\it 100}, 35.
\bibitem{moeller9046} M\"oller, D.; Fischer, J. {\it Mol. Phys.} {\bf 1990}, {\it 69}, 463.
\bibitem{moeller9214} M\"oller, D.; Fischer, J. {\it Mol. Phys.} {\bf 1992}, {\it 75},  1461.
\bibitem{lustig8817} Lustig, R. {\it Mol. Phys.} {\bf 1988}, {\it 65}, 175.
\bibitem{barker7378} Barker, J. A.; Watts, R. O. {\it Mol. Phys.} {\bf 1973}, {\it 26}, 789.
\bibitem{andersen} Andersen, H. C. {\it J. Chem. Phys.} {\bf 1980}, {\it 72}, 2384.
\bibitem{shevkunov8824} Shevkunov, S. V.; Martinovski, A. A.; Vorontsov-Velyaminov, P. N. {\it High Temp. Phys. (USSR)} {\bf 1988}, {\it 26}, 246.
\bibitem{nezbeda9139} Nezbeda, I.; Kolafa, J. {\it Mol. Sim.} {\bf 1991}, {\it 5}, 391.
\bibitem{vrabec0243} Vrabec, J.; Kettler, M.; Hasse, H. {\it Chem. Phys. Lett.} {\bf 2002}, {\it 356},  431.
\bibitem{fincham8645} Fincham, D.; Quirke, N.; Tildesley, D. J. {\it J. Chem. Phys.} {\bf 1986}, {\it 84}, 4535.
\bibitem{lotfi9213} Lotfi, A.; Vrabec, J.; Fischer, J. {\it Mol. Phys.} {\bf 1992}, {\it 76}, 1319.
\bibitem{guggenheim4525} Guggenheim, E. A. {\it J. Chem. Phys.} {\bf 1945}, {\it 13}, 253.
\bibitem{rowlinson1969} Rowlinson, J. S. {\it Liquids and Liquid Mixtures}; Butterworth: London, 1969.
\bibitem{lisal04} Lisal, M.; Aim, K.; Mecke, M.; Fischer, J. {\it Int. J. Thermophys.} {\bf 2004}, {\it 25}, 159.
\bibitem{johnson94} Johnson, K.; M\"uller, E. A.; Gubbins, K. E. {\it J. Phys. Chem.} {\bf 1994}, {\it 98}, 6413.

\bibitem{huang} Huang, Y.-L. {\it Vapor-liquid equilibria of polar mixtures by molecular simulation}, MSc Thesis, University of Stuttgart, 2005.

\bibitem{expC2H2+R152A} Lim, J. S.; Lee, Y.-W.; Kim, J.-D.; Lee, Y. Y. {\it J. Chem. Eng. Data} {\bf 1996}, {\it 41}, 1168.
\bibitem{expR142B+R113} Laugier, S.; Richon, D.; Renon, H. {\it Fluid Phase Equilib.} {\bf 1994}, {\it 93}, 297.
\bibitem{expCO2+R12} Lavrenchenko, G. K.; Nikolovsky, V. A.; Baklai, O. V. {\it Kholod. Tekh.} {\bf 1983}, {\it 6}, 41.
\bibitem{expR22+R142B} Cao W., Yu H., Wang W. {\it J. Chem. Ind. Eng. (China)} {\bf 1997}, {\it 48}, 136.
\bibitem{expPropylene+R115} Kleiber, M. {\it Fluid Phase Equilib.} {\bf 1994}, {\it 92}, 149.
 
 


\end{thebibliography}
\end{document}